# Review on physical impedance models in perovskite solar cells

Rajat Kumar Goyal[a, b], and M. Chandrasekhar[b, c*]

[a] Interdisciplinary Research Division, Indian Institute of Technology, Karwar, Jheepasani, 342030, Jodhpur, Rajasthan, India

[b] Department of Physics and Astronomy, National Institute of Technology Rourkela 769008, Odisha, India

[c] Department of Physics, CVR College of Engineering, Hyderabad 501510, Telangana, India.

∗Corresponding author
Email address: chanduphy2004@gmail.com (M. Chandrasekhar)

**Abstract:** The significant advancements in perovskite solar cell (PSC) research and development have reignited optimism for the practicality of solar energy. A deeper comprehension of the fundamental mechanisms is essential for enhancing the current solar technology. Impedance spectroscopy (IS) characterization, in conjunction with other analytical methods, stands out as the primary approach for gaining insights into the internal workings of solar cells. The continuous analysis and refinement of electrochemical spectroscopy data for PSCs have unveiled intricate operational details. Despite the ongoing improvements, the absence of a centralized repository for this information poses challenges for tracking progress effectively. This review paper aims to consolidate the physical phenomena behind the development of PSC research, aiming to establish a comprehensive knowledge base on impedance plots, equivalent circuits, and related processes.

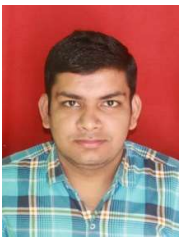
**Rajat Kumar Goyal** received his BSc in Physical sciences from the Ramjas College, University of Delhi, India, and MSc. in Physics from the Department of Physics and Astronomy, National Institute of Technology, Rourkela, India, and he is pursuing an MTech. Degree in Quantum Technology focused on Quantum materials, devices, and sensors from an interdisciplinary research platform (IDRP), Indian Institute of Technology Jodhpur, India. During His Master's, He qualified for national level exams CSIR JRF + NET and GATE. His research interests include an experimental physics perspective on the latest technology and applications, with various materials synthesis and characterizations, quantum materials, devices and sensors, and semiconductor devices. His work is in a multidisciplinary domain, and he has a background in science and technology.

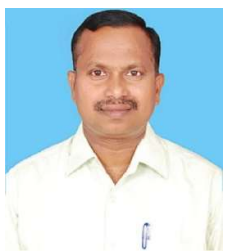
**Dr. M. Chandrasekhar** received a B.Sc. in 2002 and an M.Sc. in Physics in 2004 from Osmania University, Hyderabad, Telangana, India. He then worked as a faculty member for various degrees and engineering colleges. He qualified for APSET and GATE in 2012. He completed doctoral research in 2018 at the Department of Physics and Astronomy, National Institute of Technology, Rourkela, India. His main research interests include the synthesis of lead-free ceramics for energy storage applications, thin film preparation using spin coating, and the synthesis of various ceramic materials via solid-state reaction, sol-gel, and combustion reaction routes.

# 1. Introduction

In the dynamic landscape of photovoltaic research, perovskite solar cells (PSC) have emerged as a promising frontier, offering unparalleled efficiencies and versatility. These solar cells are based on perovskite-structured materials, typically metal halide perovskites, which exhibit excellent light-absorption properties and charge transport characteristics. The unique properties of perovskite materials make them promising candidates for next-generation solar cell technologies. Within a decade, perovskites exceptional properties and innovative architectural designs have increased the solar cell efficiency from 3.8% [1] to 26.7%.[2] Better understanding of operating and degradation mechanisms can lead to future efficiency and stability enhancements. Solar cells operate through various physical processes, including recombination rates, carrier mobilitiy, and the effectiveness of contact points in extracting charge.[3], [4] Understanding these mechanisms is crucial for optimizing their performance. The primary aim of studying solar cells is to analyze their steady-state behaviour, typically represented by the current density-voltage (J -V) curve, which dictates the power conversion efficiency (PCE) from sunlight to electricity. However, solely examining the steady state behaviour does not provide insights into the dynamic phenomena governing their performance. To unravel these dynamics, it is essential to introduce external signals and observe the system's response.

The conventional method used to assess the electrical properties of PSCs involves generating current-voltage (J-V) curves through both forward and reverse scans.[5][6][7] However, the presence of anomalous hysteresis in these measurements can compromise the accuracy of crucial parameters like short circuit current ($J_{SC}$), open circuit voltage ($V_{OC}$), and fill factor (FF). The severity of this hysteresis and the scan rate play a significant role in determining these parameters.[4], [8]–[10] Hysteresis in PSCs is attributed to various factors including ionic conduction, contact properties, and surface polarization.[11], [12] [13][14][15][16]. The complexity involved suggested, there is a need for advanced techniques capable of capturing a broader range of data to enhance the characterization of PSCs.

The most commonly employed $ABX_3$ structure in literature is $MAPbI_3$[1], which possesses a low bandgap (1.55eV) [17], high light extinction coefficient, long charge carrier diffusion length, long charge carrier lifetime, and high charge carrier mobility.[18][19][20] Factors like a high optical absorption coefficient, a long carrier diffusion length and suppressed recombination, along with a well-balanced charge transfer results in its better photovoltaic power deliverability. However, $MAPbI_3$ based PSCs exhibit low stability under environmental conditions (such as light, heat, and oxygen), ionic migration, and interface/contact issues.[21] To overcome these, various strategies such as approaching towards lower dimensions, chemical composition modifications, and surface passivation are employed.[22] Additionally PSCs exhibit unique properties uncommon in semiconductor devices, including anomalous hysteresis during J-V measurements[8][10][13] and huge switchable photocurrents[6].

The fundamental idea behind Electrochemical Impedance Spectroscopy (EIS) is to disturb a balanced electrochemical system with an electrical signal and then study how it responds. There are two main approaches: potentiodynamic EIS (PEIS) and galvanodynamic EIS (GEIS), which either perturb the system's potential or current, respectively. The system's resistance to potential disturbances is measured using alternating-current (AC) impedance, which varies with signal frequency and becomes the ohmic resistance under stationary direct-current (DC) conditions. D.D. Macdonald outlines several benefits of EIS: (1) its linearity allows for analysis using Linear Systems Theory, (2) an infinite frequency impedance spectrum provides the same insights as linear electrical techniques, (3) EIS efficiently produces a large volume of information in comparison to the collected data, and (4) data validity can be assessed using mathematical techniques, regardless of the experimental processes involved.[25] A thorough understanding of device behaviour is made possible by EIS, which also assists in the discovery of critical insights into the performance of devices.[11][26][27][28][29] EIS is able to offer crucial details such as charge recombination, dielectric variables, and interfacial effects, in addition to the position and density of traps.[14][30][31][32] This through examination not only helps to understand phenomenon such as degradation, stability and hysteresis but also enables to gain deeper insight into the underlying.[33]. In spite of the fact that current simulation technology makes the process of collecting and fitting impedance arcs with similar circuits more straight forward, difficulties arise when attempting to associate and attribute particular components of the impedance spectrum to implicit mechanisms that are present inside the PSC. Getting rid of this uncertainty is something that continues to be of interest in this subject. Figure 1 represents the systemic overview of IS.

Moreover, impedance measurements provide a comprehensive insight into a wide time spectrum, facilitating the assessment of both gradual phenomena such as degradation and ionic dispersion, alongside more rapid processes like electron transport and recombination.[34][35] Nevertheless, the interpretation of impedance spectra has sparked ongoing discussion. The intricacy arises from the concurrent operation of diverse physical mechanisms, encompassing electron and hole transport, as well as ionic migration within the device. Untangling and comprehending these interlinked phenomena present notable hurdles in accurately deciphering impedance data for Photovoltaic Solar Cells (PSCs). Establishing connections between electrical components and specific mechanisms within impedance spectra is paramount, yet often contentious. Despite the absence of an all-encompassing model spanning the entire frequency spectrum and various device configurations, Impedance Spectroscopy (IS) persists as a valuable diagnostic instrument for scrutinizing dielectric properties, internal processes, and interfacial characteristics in PSCs. Nonetheless, the interpretative nature of this characterization technique fosters debate, underscoring the necessity for meticulous experimentation and the consideration of diverse paradigms to enhance interpretation and comprehension of underlying device behavior.

This review emphasizes the significance of utilizing equivalent circuits in modelling impedance spectra to gain a deeper understanding of the physical processes inherent in perovskite solar cells (PSCs). The primary aim of this study is to elucidate the role of impedance spectroscopy (IS) measurements in unravelling the operational principles of PSCs. Initially, a concise overview of IS and fundamental Equivalent Circuit Models (ECMs) is provided, followed by an exploration of various phenomena through ECMs tailored for PSCs. This covers reliable measurement practices, data interpretation, and the challenges encountered in applying IS to PSC research. The review delves into how IS has evolved into a pivotal tool for furnishing detailed device-level insights, with the ultimate goal of enhancing the efficiency and reliability of PSCs. Furthermore, it delves into prospective research directions, pinpointing areas ripe for further exploration in this dynamically advancing domain. This paper serves as a valuable reference for researchers, scientists, and engineers looking to enhance their understanding of IS techniques and their applications in PSCs.

## 2. Understanding of IS & basic models

IS involves applying a small-amplitude stimulus (voltage or current) to a system and measuring the subsequent response (current or voltage) over several frequencies. For accurate measurements, a small-amplitude perturbation is necessary to establish a linear relationship between the signal and system response. This can be done by either two electrode setup or a three-electrode configuration. In two electrode setup, reference and counter electrodes of investigated cell are in electrical contact, it is useful for analyzing attributes in cases where it is unnecessary to distinguish between the individual electrode processes, such as in symmetric half cells. Three electrode setup configuration gives advantage to distinguish impedances from counter or working electrodes, it is preferable because it enables accurate differentiation of the interface properties related to each electrode (cathode or anode). In this section, we will cover the mathematical foundation of EIS, the general (Basic/foundation) models used for IS, validation of EIS data and different timescales.

*2.1 Mathematical formalism for impedance data*

The system can be analyzed using IS, is based on the frequency-dependent potential or current response of the system. A potential-controlled excitation Impedance Spectroscopy (PEIS) and corresponding current response can be represented as:

$$E(t) = E_0 \sin(\omega t) \qquad (1)$$

$$I(t) = I_0 \sin(\omega t + \phi) \qquad (2)$$

Where $E_0$ is low voltage amplitude, $I_0$ is current amplitude, and $\phi$ represents the phase shift. The impedance of the circuit can be defined as:

$$Z = \frac{E(t)}{I(t)} = \frac{V_0 \sin(\omega t)}{I_0 \sin(\omega t + \phi)}$$

$$= Z_0(\cos(\phi) + j\sin(\phi)) = Z_0 \exp(j\phi) = Z' + jZ'' \qquad (3)$$

Where phase $\phi = (\omega t)$ and $Z'$, $Z''$ is real and imaginary part of impedance. Which is represented in complex plane on X-axis and Y-axis, respectively. As shown in figure1. Impedance can be represented either by its amplitude and phase ($|Z|$, $\phi$) or by its real and imaginary parts (Re(Z), Im(Z)), providing insights into both the resistive and reactive elements within the circuit. This representation is valid for every frequency within the spectroscopic range, forming a three-dimensional array with frequency as a discrete variable. Typical representations of this data include Nyquist (Im(Z) vs. Re(Z)) or Bode ($|Z|$ vs. ω) plots, which offer projections along one of the three axes, allowing for visualization of impedance behaviour at different frequencies. However, these projections inherently obscure some information, which becomes significant when fitting a model to the data. For instance, Nyquist plots provide insight into the frequency response and stability of a system but lose information about magnitude. Similarly, Bode plots highlight magnitude and phase but lack direct representation of the impedance's complex nature. Therefore, careful consideration is required when selecting a representation to ensure that all relevant information is preserved, especially when employing mathematical models to analyze or predict circuit behaviour.

*2.2 Overview of EEC elements for PSCs*

Traditionally, the interpretation of IS data has relied on Equivalent Electrical Circuit (EEC) models, which represent spectra using combinations of circuit elements like resistor, capacitor, constant phase element (CPE), and inductor arranged in series or parallel configurations. While effective for simple systems with well-defined physics, this approach may not adequately distinguish between proposed local processes, such as reaction mechanisms. To effectively characterize PSCs using IS, it is essential to understand capacitances and resistances and their association with different physical processes, with capacitances and inductances often linked to phenomena like space-charge polarization and electro-crystallization processes at electrode surfaces. [36] Figure 2 illustrates the diverse impedance (R, L, C) relevant to PSC studies, extensively discussed in existing IS literature. In this section, we made a detailed discussions about EEC elements (R, L, C).

*(a) Overview of Resistance*

In impedance analysis, resistance plays a crucial role in understanding the behavior of various systems, including ionic and electronic conductivity. IS technique offers a comprehensive understanding of resistance, both in terms of frequency-dependent and frequency-independent characteristics. This analysis helps in identifying and quantifying different types of resistances present within the system, such as charge transfer resistance, charge recombination resistance, and ohmic resistance. Moreover, impedance analysis allows for the characterization of resistive behaviors under different conditions, aiding in the optimization and enhancement of system performance. Figure 2(b) illustrates the various resistances typically observed in PSCs. Mathematically, the generic representation of resistance in terms of potential difference and current can be expressed as follows:

$$R_{Tot} = \left(\frac{\partial J}{\partial V}\right)^{-1} \quad (4)$$

Where $R_{Tot}$ is total resistance of system, J is current density and V is potential difference.

**Charge transfer resistance *($R_{CT}$)* or transport resistance *($R_{TR}$)*:** It is commonly used in electrochemistry to mimic interface charge transfer phenomena like ionic current to electrical current across the electrochemical double layer. Band bending at a contact junction or grain boundary can prevent current flow at an interface, causing $R_{CT}$ which controls the charge transfer at these interfaces and is affected by voltage and temperature. $R_{CT}$ is essential for efficient charge collection or extraction in PSCs at the interfaces between the perovskite layer and the electron and hole transport layers (ETL and HTL, respectively). $R_{CT}$ estimates the energy needed (or increased) to transfer mobile charges over a substrate and can be described in terms of material properties. It is inversely proportional to substance conductivity and mathematically stated as: [37]

$$R_{TR} = d.\sigma^{-1} = d.(z.q.n.\mu)^{-1} \qquad (5)$$

where d is the sample thickness or conduction channel length, z is the number of charges per carrier (e.g., electron, hole, or ion, and for PSCs z = 1), q is the elementary charge, n is charge carrier concentration, and m is carrier mobility. $R_{TR}$ is typically detected at high frequencies (~100 kHz) in PSCs.[38]

**Dynamic resistance ($R_d$):** Dynamic resistance of a solar cell's changes with the change in voltage and current. This is the derivative of voltage versus current. In the solar cell, dynamic models, parasitic capacitance and dynamic resistance $R_d$ are commonly linked. Existing literature has not extensively explored the dynamic resistance in PSCs. Previous study has ignored this element, despite its potential importance in explaining PSC behaviour. [39]

**Recombination resistance ($R_{rec}$):** Resistance occurs when an electron (or hole) moves from the conduction (or valence) band to a bandgap defect or surface state. Conventionally, $R_{rec}$ is measured at low frequencies (<100 Hz), but practically, it is measured in EIS circuits at frequencies above 100 kHz.[40], [41] $R_{rec}$ in perovskites is affected by carrier density, which is altered by voltage, light, and temperature. Therefore, $R_{rec}$ usually depends on carrier density (n) and temperature (T). Transport resistance ($R_{TR}$) increases as carrier density falls due to recombination. This links $R_{rec}$ with transport resistance.[37] Quantifying such linked processes with IS is difficult. A mathematical expression for $R_{rec}$ is:

$$R_{\text{rec}} == \frac{mk_bT}{qj_o} \frac{e^{-qV}}{nk_bT} \qquad (6)$$

**Series resistance ($R_s$):** It is recommended that it to be kept low because it frequently originates from device contacts. ITO/FTO, which stands for transparent conductive oxide windows, are the cause of this phenomenon in thin film PSCs. ).[37] It corresponds to a shift along the real axis away from the origin, determining $R_s$ in the Nyquist plot is an easy process.[42] The presence of sheet resistance of contact materials can also be verified by the use of the Nyquist plot, which is a measurement technique.

*(b) Overview of capacitance*

Capacitance is essential in the impedance spectroscopy, enabling the characterization of material electrical properties. It provides insights into charge transfer kinetics, diffusion processes, and double-layer formation at interfaces. Figure 2(c) illustrates the various capacitance typically observed in PSCs.

**Chemical capacitance or diffusion capacitance ($C_\mu$):** It has significant concepts in electrochemistry and semiconductor physics. Chemical capacitance refers to the charge storage capacity related to chemical reactions at the interfaces, seen in electrochemical systems like lipid bilayers.[41] Conversely, diffusion capacitance arises from charge carrier transport between terminals, notably in semiconductor devices such as P-N junction diodes. It results from charge redistribution under an electric field, forming a depleted region akin to a capacitor. $C_\mu$ quantifies the impact of displacing $E_{Fn}$ on the electron or hole density (n), as defined by the equation:

$$C_\mu = q^2 \frac{\partial n}{\partial E_{Fn}} \qquad (7)$$

Here, q refers to the charge, and the equation accounts for a similar relationship for holes [47]. This measure is typically observed in the low frequency range of 0.1-10 Hz [48], [49]. Here n is carrier density within system and the carrier density follows Boltzmann statistics, it can be written as:

$$n = \frac{n_0 e^{(E_{Fn}-E_{F0})}}{k_b T} \qquad (8)$$

Where, $n_o$ represents a constant, $E_{Fn}$ is the Fermi energy, $E_{FO}$ is the reference Fermi energy, $K_b$ is the Boltzmann constant, and T is the temperature. If we put equation 8 into equation 7 we get

$$C_\mu = \frac{q^2 n}{k_b T} \qquad (9)$$

The total chemical capacitance is directly proportional to the thickness of the film, expressed as:[43]

$$C_{\mu t} = c_\mu d \qquad (10)$$

In forward bias diodes, diffusion capacitance is due to the accumulation of minority charge carriers at the surface/boundary of the depletion region. This capacitance is generally observed in mid-high frequency range, typically above kilohertz (>1kHz). The diffusion capacitance can be approximated as:

$$C_{diff} \sim \frac{qnd}{kT} \exp\left(\frac{qV}{kT}\right) \quad (11)$$

Here, $V$ represents the voltage determining the width of the depletion region, $n$ is the density of minority carriers, and $T$ is the temperature. In solar cells, the diffusion capacitance is considered in parallel to the drift transport.

**Geometric capacitance**: It is intimately linked to dielectric relaxation and plays a vital role in charge separation and polarization in an electric field. The geometrical capacitance is determined by the instantaneous relaxation resulting from the displacement of holes and electrons from their equilibrium positions.[44] This phenomenon observed when the dielectric material reacts to the applied electrical field, leading to the creation of quantifiable capacitance. Studying the connection among charge separation, polarization, and dielectric capacitance offers useful insights into the behaviour of electrical systems in different applications, especially at high frequencies ($< 10^5$ Hz). Geometrical capacitance is defined as:

$$C_g = \frac{\varepsilon_0 \, \varepsilon_\infty A}{d} \quad (12)$$

Where $\varepsilon_0$ dielectric constant, $\varepsilon_\infty$ dielectric constant at high frequency, A is area and d is thickness of material.

**Electrode capacitance:** Electrode polarization is a phenomenon that takes place in ionic conducting systems because ions are unable to travel through the metal collector contact. A decrease in conductivity is brought about at low frequencies as a result of this polarization. Within the contact interface, the concentration of ions results in the production of a capacitance that is connected to the surface space charge. It is possible for this capacitance to exist close to the contact if the film is sufficiently thick and can continue to exist regardless of the thickness of the film (d). The polarization of the electrode results in a significant rise in the values of the apparent permittivity. This is a characteristic that is shared by a multitude of ionic systems. It is described as the capacitance that is related with electrode polarization, which happens within a particular frequency range.

$$C_{sc} = \frac{\varepsilon\varepsilon_0}{\lambda_D} \quad (13)$$

Here, $\lambda_D$ represents the Debye length, defined as

$$\lambda_D = \sqrt{\frac{\varepsilon\varepsilon_0 K_b T}{q^2 N}} \quad (14)$$

**Contact capacitance:** A work function offset at a metal semiconductor interface results in the creation of a Schottky barrier. The barrier causes depletion zones to form at the semiconductor-metal interface, leading to the generation of capacitive responses. The capacitive response magnitude is dictated by the width of the depletion layer (w). By adjusting the width of the depletion layer, it is possible to influence the capacitive characteristics of the system. Studying these capacitive reactions is essential for designing and optimizing electrical devices like diodes and transistors.[44] Contact capacitance is typically detected at frequencies above 100 kHz and can be characterized mathematically as:

$$C_{dl} = \frac{\varepsilon_0 \varepsilon_\infty}{w} \quad (15)$$

*(C) Overview of constant phase element*

Constant phase element (CPE) is an equivalent electrical circuit component used in modelling the behaviour of systems where the impedance does not follow ideal capacitor characteristics. It is particularly relevant in the context of EIS, where real systems exhibit non-ideal behaviour that deviates from a perfect capacitor. The CPE is a versatile element that can represent various phenomena, such as imperfect capacitors, double-layer capacitance in electrochemical systems, and other systems with complex impedance responses.

The CPE is used to model the behaviour of systems with non-ideal impedance responses, often resembling a depressed semicircle in Nyquist plots. Mathematically, the impedance of a CPE is expressed as

$$Z_{CPE} = \frac{1}{Q_0 (j\omega)^n} \quad (16)$$

Where $Q_0$ is admittance at $\omega = 1$ rad/s and n is a constant phase angle between 0 to 1. The CPE can represent systems with varying degrees of imperfection, from ideal capacitors (n = 1) to pure resistors (n = 0), with values in between indicating different levels of deviation from ideal behaviour. The CPE is commonly used in electrochemical systems to model phenomena like double-layer capacitance, rough surfaces, and heterogeneous reactions, providing a more accurate representation of the system's electrical properties.[45]

*(D) Overview of Inductance*

The inductance prospective of PSCs is a topic that has garnered significant attention in recent research. Studies have highlighted the presence of inductive effects in perovskite solar cells, which are typically associated with the dynamic hysteresis effects observed in the current density-voltage characteristics of solar cells. [46] These effects are primarily attributed to ionic migration within the solar cells, including the kinetic effects of ions causing voltage delay and charge accumulation [47][48], phase-delayed recombination[49], electron injection effects[50][51], and ionic-mediated recombination [52] influencing the behaviour of the cells.

Bisquert J. and Guerrero A. discovered a 'chemical inductor' in an electrochemical system exhibiting inductive behaviour without an electromagnetic origin. Chemical inductors dynamically affect physiochemical processes with their two-dimensional structure comprising a slowing-down element and a fast conduction mode. This chemical inductor causes negative voltage transient spikes, inverted J-V curve hysteresis, and cyclic voltammetry. [53] To understand this phenomenon, Gonzales et al. built a neuron-style model using differential equations that can follow the change from capacitive to inductive behaviour using IS and J-V measurements with different sweep rates. [53] This model illuminates low-frequency information, notably about ion-controlled surface recombination processes.

It is the surface polarization model for PSCs that provides the basis for the inductor element in ECs, which is the physical genesis of the element .[12] The autonomous relaxation of the internal voltage at the contact surface in a PSC is described by this model, which also describes the phenomenon to be described. The relaxation equation can be expressed as follows:

$$\frac{dV_s}{dt} = \frac{-V_S-(V-V_{bi})}{\tau_{kin}} \qquad (17)$$

Here, $V_s$ is surface potential $\tau_{kin}$ is relaxation kinetic constant, $V$ is total voltage between the contacts, and $V_{bi}$ is the built-in voltage. The value of $\tau_{kin}$ is determined by the speed at which ions move during surface polarization and depolarization induced through external bias.[54]

The phenomenon of inductive behaviour in PSCs can be explained by a set of kinetic equations that describes the relaxation of an internal slow state variable influenced by an external "fast" variable. These mechanism has been employed in the field of electrochemistry, [55],[56], [57] and more recently, it has been extended to encompass the impedance of memristors and biological neurons.[58] The dynamic model containing the chemical inductor follows the following kinetic equations:

$$\tau_m \frac{dV}{dt} = f(V) + R_I(-w + I) \qquad (18)$$

$$\tau_k \frac{dw}{dt} = \frac{1}{R_a} V - w \qquad (19)$$

Where $V$ is external voltage, $I$ is current across the device, $w$ is internal current that shows the slow transient effects (such as ionic effects), [44] $\tau_k$ is recovery current response time, $R_I$ is the channel resistance, $R_a$ is the recovery current resistor, $f(V)$ is a conductivity (function of main channel), and $w$ is the slow relaxation process. The useful parameters derived from these equations are:

$$\text{Capacitance } C_m = \frac{\tau_m}{R_1} \qquad (20)$$

$$\text{Resistance } R_b = -\frac{R_1}{f'} \qquad (21)$$

$$\text{Inductance } L_a = \tau_k R_a \qquad (22)$$

$$\text{Impedance } Z(\omega) = [R_b^{-1} + C_m s + (R_a + L_a s)^{-1}]^{-1} \qquad (23)$$

The use of series RL branching is gaining preferences to fit the EIS spectra, which provides important information connected to underlying physical processes. This is because the usage of series RL branching is a result of rigorous derivation from fundamental physical laws that relate to the impedance structure of a slow/fast dynamical system of PSCs. [48], [59]

*2.3 Understanding of simulated basic EEC models*

Understanding of basic models involves comprehending the fundamental principles of modelling electrochemical systems using equivalent circuits. The purpose of using equivalent circuit models in EIS is to analyze and interpret the complex impedance data obtained from electrochemical systems such as batteries, by representing the internal behaviour of the system through external characteristics. These models serve as a representation of the internal behaviour of the system through external characteristics, allowing researchers to understand the underlying electrochemical processes. EEC models typically consist of various elements like resistors, capacitors, inductors, and specialized electrochemical elements, such as Warburg diffusion elements, to simulate the behaviour of the system accurately. The process of creating an EEC model starts with predicting the system elements that contribute to impedance, followed by building an equivalent circuit model incorporating these elements in logical series and parallel combinations. Each element in the model has a known impedance behaviour described by specific parameters. By utilizing tools like Zview, Zplot, Zsimpwin, EIS analyzer and others, researchers can construct these models to analyze EIS spectra and gain insights into the physical processes within electrochemical cells. Understanding of basic EEC models is essential for interpreting EIS data, predicting system behaviour, and guiding decision-making in electrochemical research and development so in this section we will discuss building block circuits of EEC models.

In Figure 3 and Figure 4, Nyquist, Bode magnitude and phase plot are represented for some basic circuits. In this we will explain each one in very brief with impedance equation. In Figure 3(a), only a resistor is present, for this impedance equation is $Z = R_0 + j.0$. for this we can see from plot real part is R and imaginary part is 0 so Nyquist plot signifies a single point on real axis. It represents a constant value of Impedance (which is equal to R) for all frequencies. Bode magnitude plot and Bode phase plots shows a straight line.

$$Z = R_0 + j.0 = Z' + jZ'' \Rightarrow Z' = R_0, Z'' = 0 \quad (24)$$

$$|Z| = \sqrt{(Z')^2 + (Z'')^2} = \sqrt{(R_0)^2 + (0)^2} = R_0 \quad (25)$$

$$\phi = \tan^{-1}\left(\frac{Z''}{Z'}\right) = 0 \quad (26)$$

Similarly, if we observe only when capacitance is present in the circuit as shown in Figure 4(b). Equation of impedance

$$Z = 0 + \frac{1}{j\omega C} = 0 - j\left(\frac{1}{\omega C_1}\right) = Z' + jZ'' \Rightarrow Z' = 0, Z'' = -\frac{1}{\omega C_1} \quad (27)$$

$$|Z| = \sqrt{(0)^2 + \left(\frac{1}{\omega C_1}\right)^2} = \frac{1}{\omega C_1} \quad (28)$$

$$\phi = \tan^{-1}\left(\frac{Z''}{Z'}\right) = -\frac{\pi}{2} \quad (29)$$

From this calculation it is clear that $Z''$ is inversely proportional to frequency and capacitance. Similar observation we see in Figure 3(b), in Nyquist plot a straight line along y-axis, Bode magnitude plot shows a straight line with slope -1 and Bode phase plot shows a straight line at $\phi = -\pi/2$. Here we have to note this point that for ideal capacitor phase different is π/2 but for real capacitance it is not equal to π/2, generally it is less than π/2 so we use a constant phase element (CPE).

When only inductance is present in the circuit as shown in Figure 3(c). equation of impedance

$$Z = 0 + j\omega L = Z' + jZ'' \Rightarrow Z' = 0, Z'' = \omega L, \quad |Z| = \sqrt{(0)^2 + (\omega L)^2}$$

$$= \omega L, \quad \phi = \tan^{-1}\left(\frac{Z''}{Z'}\right) = \frac{\pi}{2} \quad (30)$$

Real part of impedance is zero and imaginary part is proportional to frequency and inductance. So, Nyquist plot shows a straight line along -y axis as in this case Phase difference is $\phi = \pi/2$ so in Bode phase plot we observed a straight line parallel to x-axis and in Bode phase plot we observed a straight line with slope 1. As represented in Figure 3(c).

When resistance and capacitance is present in the circuit as shown in Figure 3(d). equation of impedance

$$Z = R_0 + \frac{1}{j\omega C} = R_0 - j\left(\frac{1}{\omega C_1}\right) = Z' + jZ'' \Rightarrow Z' = R_0, Z'' = -\frac{1}{\omega C_1} \quad (31)$$

If we see the pattern in Nyquist plot (Figure 3(d)), it a combination of Nyquist plot for Only capacitance (Figure 3(b))and only resistance (Figure 3(a)). If we observe real and imaginary part at higher and lower frequency region, at high frequency region imaginary part is neglected and real part dominates similarly in lower frequency imaginary part dominates. So, we observe (Figure 3(d)) a straight line (independent of frequency) in higher frequency region. In lower frequency region, imaginary part increase at a specific frequency, ω = 1/R₀C when real part is equal to imaginary part, after this response is dominated by imaginary part. This transition we can also observed by S-shaped curve in Bode phase plot.

When the circuit contains a capacitance and resistance with parallel connection, Figure 4(a), the equation of impedance is

$$Z(\omega) = \frac{1}{\frac{1}{R_1} + jwC_1} = \frac{R_1}{1 + j\omega R_1 C_1} = \frac{R_1}{1 + (\omega R_1 C_1)^2} - j\frac{\omega R_1^2 C_1}{1 + (\omega R_1 C_1)^2} \quad (32)$$

It is observed a semicircle in Nyquist plot. In high frequency case, capacitive reactance tends to zero and so it works as a short circuit and all current passes through it. In lower frequency case, capacitive reactance tends to infinity so all current passes through the resistor, so impedance contain only real part ($R_1$). From High to low frequency change, there is a single characteristic frequency (Impedance part is maximum) corresponding to equal value of reactance and resistance.

$$\frac{1}{\omega C_1} = R_1 \Rightarrow \omega_{Z''max} = \frac{1}{\tau} = \frac{1}{R_1 C_1} \quad (33)$$

Here $\tau = R_1 C_1$ is the characteristic time constant of the system. At this frequency real and imaginary part becomes equal. Bode mode plot 4(a), according to above equations at very low frequencies $|Z| = R_1$, while at very high frequencies $|Z| = 1/\omega C_1$. The slope of the curve is changing it is represented in the Figure 4(a) by a dotted line circle around the frequency corresponding to the time constant of the system.

If we connect a $R_0$ resistance in Figure 4(a) than circuit looks like Figure 4(b) and impedance is represented by

$$Z(\omega) = R_0 + \frac{R_1}{1 + (\omega R_1 C_1)^2} - j\frac{\omega R_1^2 C_1}{1 + (\omega R_1 C_1)^2} \quad (34)$$

At both very high and very low frequencies, the circuit responds resistively, shifting the semicircle in the Nyquist plot to the real axis. This shift occurs at the same value as the circuit's ohmic resistance ($R_0$) as represented in Figure 4(b). At very high frequencies, where ω tends to infinity and the capacitive reactance ($X_C$) approaches zero, the impedance (Z) simplifies to R0. Conversely, at very low frequencies, as ω approaches zero and $X_C$ tends to infinity, Z becomes $R_0 + R_1$. These boundary conditions manifest as the first and second crossing points of the semicircle on the real impedance axis in the Nyquist plot. Unlike a simpler circuit (Figure 9A), this circuit exhibits two-time constants, leading to two breakpoints in the Bode magnitude plot. The first time constant ($\tau_1$), occurring in the high-frequency domain, is affected by $R_0$ and can be computed as $\tau_1 = [(R_0 R_1)/(R_0 + R_1)]C_1$. The second time constant ($\tau_2$), which represents the slower response, can be derived from the Nyquist plot similar to a simple RC parallel circuit, where $\omega_{Z''max} = \frac{1}{\tau} = \frac{1}{R_1 C_1}$. Examining the Bode plot of this circuit, the variation in |Z| and phase across a broad frequency spectrum follows an S-shaped and a bell-shaped curve, respectively. These curves elucidate the change in impedance magnitude and phase angle with frequency.

If we add a parallel RC circuit into Figure 4(b), the circuit is represented in Figure 4(c). The equation of impedance

$$Z(\omega) = R_0 + \left[\frac{R_1}{1 + (\omega R_1 C_2)^2} + \frac{R_2}{1 + (\omega R_2 C_1)^2}\right] - j\left[\frac{\omega R_1^2 C_2}{1 + (\omega R_1 C_2)^2} + \frac{\omega R_2^2 C_1}{1 + (\omega R_2 C_1)^2}\right] \quad (35)$$

The circuit in Figure 4(c), which is analogous to the analysis presented earlier Figure 4(b), demonstrates two distinct time constants, namely $\tau_1$ and $\tau_2$, which correspond to the maximum value of $\omega_{Z''_{max}}$ at each semicircle. Depending on the ratio of the values of $\tau_1$ and $\tau_2$, the Nyquist plot exhibits either two semicircles that are well-resolved (where $\tau_1$ is greater than 100 $\tau_2$) or poorly resolved (where $\tau_1$ is less than 100 $\tau_2$). However, when $\tau_1$ equals $\tau_2$, a single semicircle is observed. Additionally, the pattern of the Bode magnitude and phase charts is influenced by the ratio between the $\tau_1$ and $\tau_2$ values. The inclusion of a second step and peak in the plots that are produced as a result of the second time constant occurs when the value of $\tau_1$ is equal to or greater than 100 times $\tau_2$.

## 2.4 Validation of Impedance Spectra

Accuracy and validity of the impedance spectrum are crucial for the reliability and reproducibility of the results. Therefore, validation of impedance spectra is an essential step in impedance spectroscopy. Validation of impedance spectra involves checking the quality and reliability of the measured data and ensuring that they meet the necessary conditions for the analysis. Figure 5 emphasizes the validation points of IS to ensure accurate interpretation and extraction of significant data. In this we will discuss in brief about the validation steps of IS.

**Linearity and stationarity** as shown in Fig. 5(1) (pseudo) linear response means that the current follows the potential and follows Ohm's law in that area. It is easy to get linearity by picking an excitation amplitude that is small enough. Stationarity, on the other hand, is a problem that needs careful thought when measuring IS because it cannot be achieved by just setting the conditions of the experiment. If the system is not stationary, the current answer will have an extra term as

$$I(t) = I_0 \sin(\omega t + \emptyset) + I_{non-stationar}\ (t) \qquad (36)$$

Usually, impedance analysis doesn't take this extra term into account, so it will be hard to get both qualitative and quantitative results. The Kramers–Kronig analysis (KK check), on the other hand, can help show when measurement settings are not stable.

**Integrity and K-K relations :** The accuracy, dependability, and uniformity of impedance data are all parts of its integrity. It is very important to keep impedance readings free of errors, artefacts, or interference that might affect the data quality. The accuracy of impedance data can be affected by many things, such as the quality of the instruments used, the right way to calibrate them, the way the data is processed, and how noise or electrical interference is reduced. To check and keep the accuracy of impedance data, regular quality control and validation steps should be put in place.

The real and imaginary parts of a complex impedance function are mathematically linked by the K-K relations in IS. These connections come from the basic ideas of physics, like causality and analyticity. An integral transform links the real part of the impedance spectrum (Z') to the imaginary part (Z") in IS. This is what the K-K relations say. To be more specific, the K-K transform can be written as follows:

$$Z'(\omega) = \left(\frac{2}{\pi}\right) P \int \left[\frac{Z''(v)}{(v - \omega)}\right] dv \qquad (37)$$

In this case, Z'(ω) and Z"(v) show the real and imaginary parts of the impedance as a function of frequency ω and the variable of integration v. The main number of the integral is shown by the symbol P. Using the K-K transform, you can also get the imaginary part Z"(ω) from the real part Z'(v):

$$Z''(\omega) = -\left(\frac{2}{\pi}\right) P \int \left[\frac{Z'(v)}{(v - \omega)}\right] dv \qquad (38)$$

The K-K relations serve as a powerful tool for analyzing impedance data and can be utilized to validate the consistency and accuracy of experimental measurements. By leveraging the interdependence of the real and imaginary components, these relations assist in ensuring the reliability of impedance data and facilitate the extraction of meaningful information about the system being investigated.

**Stability**: The IS stability refers to the consistency and repeatability of impedance measurements over time and under varying experimental conditions. A stable IS method should yield similar results when measurements are taken repeatedly or when experimental parameters, such as temperature, humidity, or applied voltage, are altered. Stability is a critical consideration in fields like electrochemistry, materials science, and biology, where even minor changes in experimental conditions can have a significant impact on the measured impedance. To ensure stability in IS, it is crucial to carefully control and monitor the experimental conditions, including factors such as temperature, humidity, and the purity of the materials being measured. Furthermore, conducting measurements across a range of frequencies and employing appropriate data analysis techniques to eliminate noise and other sources of error may also be necessary.

**Causality and finiteness:** According to this condition there should be an exclusive cause to effect relationship between the input and output signal response of the system. Or we can say system's response should be entirely dependent on the applied perturbation. The causality of IS pertains to the meaningful physical interpretation of impedance data in relation to the underlying electrical properties of the system being measured. the causality of impedance spectra can be utilized to determine the rate of charge transfer at an electrode-electrolyte interface and identify any non-ideal behavior in the system, such as the presence of surface adsorption or charge accumulation.

The impedance of a finite IS system does not approach infinity or 0 at any frequency. This characteristic allows reliable measurement and interpretation of impedance data across a wide frequency range and ensures well-behaved mathematical models. Practically, a finite IS system means its impedance cannot have poles or zeros in the right half of the complex plane. Such behaviour would provide unbounded impedance at certain frequencies. Thus, the imaginary impedance must always be positive or zero and the real component bound at all frequencies. Impedance spectrum finiteness is important in electrochemistry because non-finite behaviour might imply electrode deterioration, system non-uniformity, or non-ideal features. Impedance spectra must be measured and analyzed accurately, therefore mathematical properties and physical interpretation must be considered.

*2.5 Classification of Relaxation Processes in the Impedance Spectra*

Relaxation processes observed in impedance spectra can be systematically classified based on their characteristic frequency response and behavior, aiding in the interpretation of electrochemical systems. Figure 6 illustrate the different dynamic processes associated with distinct relaxation times. The relaxation processes in impedance spectra can be classified into resistive-capacitive and resistive-inductive distribution of relaxation times functions, depending on the nature of the system under study. The relaxation processes can be attributed to various physical phenomena, such as charge transfer, diffusion, and ionic migration, which can affect the performance and stability of electrochemical systems.

The classification can be approached by considering the types of relaxation processes that contribute to the impedance spectra and how they are represented in equivalent circuit models. One prominent classification method involves analyzing the shape of the Nyquist plot, a graphical representation of impedance data in the complex plane. Firstly, Debye relaxation processes manifest as semicircles in Nyquist plots, signifying ideal capacitive behavior with a single relaxation time constant, often seen in simple capacitor charging and discharging phenomena. Conversely, Warburg impedance is characterized by a linear trend with a 45-degree slope in the Nyquist plot, indicative of diffusion-controlled processes prevalent in systems governed by mass transport limitations, such as porous electrodes. Additionally, the CPE is frequently employed to model non-ideal capacitive behavior, represented by depressed arcs or straight lines at lower frequencies in the Nyquist plot, particularly in systems displaying fractal or heterogeneous characteristics. Furthermore, some systems exhibit multiple relaxation processes, leading to more intricate impedance spectra with multiple arcs or peaks in the Nyquist plot, each corresponding to distinct time constants and physical phenomena, such as diverse electrode reactions or charge transfer mechanisms. Through careful analysis of impedance spectra, we can identify and classify these relaxation processes, facilitating a deeper understanding of the underlying mechanisms and behavior of electrochemical systems.

In regards of PSCs, PSCs' IS response below 100 Hz reveals cations' and anions' ionic mobility. This motion might involve bulk and contact interfacial mechanisms at different length scales.[47][50] Ionic mobility in perovskite material mostly includes charged species moving through the lattice. Temperature, defect concentration, and electric fields affect its motion. Ion mobility impacts conductivity, carrier recombination dynamics, and charge transfer. PSCs have absorber, electron, hole transport, and electrode interface layers. Ionic mobility affects device performance at certain interfaces. The perovskite/electrode interface [61] and perovskite/transparent layer interface[62] are the two most important interfaces in PSCs. PSCs undergo trap-assisted recombination at higher frequencies above (>$10^4$ Hz).[63] This technique captures electrons or holes via localized defects and recombines them with opposing carriers. These events create more PSC semicircles or loops at higher frequencies. Analyzing and fitting these IS spectra can reveal trap densities,[64] trap states,[65] energy levels,[66] and their effects on charge carrier recombination.[67], [68]

IS utilizes resistance (R) and capacitance (C) to depict relaxation processes, which involve electrical energy dissipation (Ø = 0°) and storage (Ø = 90°) in the excitation signal. Simple RC circuits with parallel R and C can explain this behaviour. The characteristic time constant ($\tau$) of this circuit is $\tau$ = RC, indicating the time it takes for the system to reach equilibrium after stimulation or perturbation. It measures physical relaxation time. Complete ideal relaxation rarely occurs, real samples often have impedance responses that differ from the ideal relaxation process. .[69] Measurements of semiconductor devices and electrochemical systems like batteries show resistive, capacitive, and inductive components. These behaviours result from polarization, electrical and ionic transportation, interfacial charge transfer, chemical processes, double layer creation, and charge recombination.[70][71][72][73] Dispersive relaxation processes, like charge distribution on an inhomogeneous electrode surface, have a distribution of time constants.[73] The overlap of relaxation processes in Nyquist spectra distorts general spectra.[74]

## 3. Physical EIS modelling for PSCs

Equivalent Circuit Modelling (ECM) is a popular IS analysis tool that helps researchers understand complicated systems. However, successful EIS analysis requires assigning significant physical components to electrical elements to build the device model, which yields consistent and trustworthy results. Assigning elements for an EEC becomes increasingly complex with the challenge of evaluating time constants for a dataset and the difficulty of decoding physical features from IS spectra. The Distribution of Relaxation Times (DRT) technique can help by transforming EIS data into a time domain, revealing time constants. However, DRT analysis is prone to statistical deviations, limiting its quantitative use. Additionally, finding the best model for EECs is not straightforward. Determining a suitable EEC involves either using an available model or deriving a physical model. While models are empirical and offer reasonably accurate fitting, verifying them is challenging. In contrast, physical models assign elements based on physical origins, providing insights into electrochemistry and charge transfer mechanisms. These models are verifiable and can be tested using other techniques or simulations. For instance, while a semi-circle in impedance spectra can be fitted with multiple RC-elements, introducing additional parameters must be weighed against fit quality. However, fitting problems, such as overfitting, must be avoided. Extra parameters should only be added if they significantly improve fit quality, considering the potential for errors in each fitting parameter.

In real life, things are more complicated than in ideal EEC. So, to make a better model for EECs, we need to consider how physical things work. This means looking at real data and the processes happening in the system we are interested in, based on what we see in experiments. In this part, we will talk about ideas and different types of models for interpreting the spectra of PSCs. Each type of model we discuss will explain specific parts of a circuit, how they are connected, and what they mean in terms of what is happening in PSCs. We will also look at how different parts of the circuit affect different frequencies, like low, medium, and high frequencies.

**Type 1. Voight circuit:**

The Voight Circuit is one of the circuits that is used in literature the most frequently over the years. As shown in Figure 7, it has become the most common EC that is utilized for the purpose of fitting the IS spectra of PSC measurements. Among the important components that are included in this circuit are $R_s$ (series resistance), $R_{ct}$ (charge transfer resistance) or $R_{tr}$ (charge transport resistance), and $R_{rec}$ (recombination resistance). These components are connected in parallel with their corresponding constant phase elements (CPE1 and CPE2). It gives information about these elements.

There are a lot of studies that are available in the body of previous literature. We took a study for analyzing the IS by Bhaskar Parida, which is illustrated in Figure 7 (a), and corresponding their circuit and impedance of circuit.[75] Bhaskar Parida investigated charge transport and recombination in PSCs by conducting IS tests on devices with mp-TiO$_2$ and c-TiO$_2$/m-TiO$_2$ layers. The findings of the study demonstrated that the c-TiO$_2$/m-TiO$_2$ layer displayed lower values for $R_s$ and $R_{tr}$ in comparison to the mp-TiO$_2$ layer under investigation. This observation demonstrates that the c-TiO$_2$/m-TiO$_2$ layer is an effective charge transport layer, and it also underlines the impact that this layer has on recombination rates. It was determined that the interface between the c-TiO$_2$/m-TiO$_2$ layer and the ETL and the Perovskite layer was responsible for the increased charge transport capabilities of the c-TiO$_2$/m-TiO$_2$ layer. The rate of charge recombination has been successfully lowered as a result of their successful efforts. As a consequence of this, the c-TiO$_2$/m-TiO$_2$ layer makes a significant contribution to the optimization of charge transport and the reduction of recombination, which ultimately results in an improvement in the overall efficiency and functionality of the PSCs.

With the use of the Voigt circuit, the researchers were able to determine the charge collecting efficiency (CCE) of PSCs by applying the equation that is presented below:[76]

$$CCE \approx \left(1 - \frac{R_{rec}(OC)}{R_{rec}(NOC)} \times 100\right) \qquad (39)$$

The symbol R$_{rec}$ denotes the recombination resistance in high frequency region, whereas OC and NOC stand for Open Circuit and Non-Open Circuit conditions, respectively.

The application of the Voigt circuit method for extracting high-frequency resistance in PSCs has been extensively reported in the literature, as evidenced by references such as [77], [78] to extract the high frequency resistance in PSCs. In order to contribute to the development of PSCs technology, the purpose of the study was to obtain insights into the recombination behaviour and CCE of PSCs. This was accomplished by the utilization of the Voigt circuit and the equations that were described earlier.

However, it should be emphasized that the Voigt circuit has difficulty comprehending certain elements of electrochemical systems, particularly when dealing with interfaces and analyzing charge transfer kinetics at the interfaces. The Voigt circuit's components are essentially resistors and capacitors/CPEs, which may not be adequate to provide a complete physical interpretation of impedance spectra when inductive effects (negative loops) are found in Nyquist plots. Furthermore, the Voigt circuit assumes a constant relationship between impedance and frequency, which may not correctly reflect the frequency-dependent behaviour commonly observed in various types of PSCs. This shortcoming further limits its usefulness for interpreting all forms of PSCs. To address these restrictions, more complex techniques, such as the Modified Randles circuit, have been used. The Randles circuit adds new features to better reflect interfacial effects, analyze charge transfer kinetics, and capture the intricacies of electrochemical systems. This method yields a more accurate and thorough interpretation of impedance data in a broader variety of PSCs.

**Type 2. Randles circuit:**

The Randles circuit is a widely used equivalent circuit model in the field of electrochemistry and electrochemical impedance spectroscopy (EIS). It is a model that is frequently utilised for the purpose of researching ion migration as well as oxidation and reduction processes. It is used to model a semi-infinite diffusion-controlled faradaic reaction to a planar electrode, and it includes the ionic resistance ($R_i$), double layer capacitance ($C_{dl}$), charge transfer resistance ($R_{ct}$), and Warburg impedance ($Z_W$) elements. The Randles circuit is arranged in a specific way to represent the physical processes occurring at the electrode-electrolyte interface, and it produces a distinctive Nyquist plot with a semicircle followed by a 45° line.

Figure 8 (a) illustrates the Randles circuit. At low frequency, the reactance of $C_{dl}$ is very high so all the current passes through the $R_{ct}$, $R_s$, and $Z_W$ corresponding circuit is represented in Figure 8(c) and IS in Figure 8(e). At higher frequencies, $R_{ct}$ is much bigger than $Z_W$ so current flow is implemented by $R_s$, $R_{ct}$, $C_{dl}$. Corresponding to this EEC is represented in Figure 8(b) and IS in Figure 8(d). in real system, we observe both pattern (semi-circle and straight line) of the Nyquist plot over a wide range of frequencies. And whole system is controlled by all parameters corresponding to this EEC is represented in Figure 8(a) and IS in Figure 8(f). If we fix all the parameters except $C_{dl}$, by simulation we can see that distortion increase with along with $C_{dl}$.

A very nice paper (for better understanding of Randles circuit) By Yao Liu examined the Effects of ionic functional groups on ion transport at perovskite interfaces.[79] For this they used device as shown in Figure 9(a,b) and corresponding EEC in Figure 9(c) and IS in Figure 9(d). if we analyzed the spectra, we get high frequency semicircle (closest to the origin) mostly reflects electronic transport, whereas the low frequency regime resulted from $MA^+$ ion transport. IS response of each device was symbolized by an initial high frequency semicircle, which was displayed by all of the devices. At low frequency, for devices with cationic PVBT-TMA and anionic PVBT-SO3 HTLs (Figure 9 d), impedance is represented by linear semi-infinite ion diffusion. EEC was utilized in order to model IS, as shown in the Figure 9(c), this system combines high frequency charge transfer with low frequency ionic transport. Here $R_s$ is used to represent the contact resistance between layers as well as in the electrode connections. The impedance at high frequencies is represented by a coupled bulk ionic and electronic charge transport/recombination resistance ($R_{tr/rec}$. Additionally, the charge stored in the perovskite is depicted, in parallel, with a chemical capacitance element ($C_\mu$). The electronic transport that is not connected with the ionic transport is represented by a parallel resistor, which is abbreviated as $R_{elec}$. Due to the fact that this low frequency feature is dependent on both the size of the ions and the temperature, it was determined that an ion diffusion component has to be incorporate into the model. As a result, the ion diffusion was modelled using a constant phase Warburg element ($W_s$) that was modified in series with an interfacial charge transfer resistance ($R_{CT}$), just like in a well-established Randles circuit. In order to mimic the accumulation of electronic and ionic charges at the interface, a double-layer capacitance ($C_{DL}$) is utilized. Devices with zwitterionic PVBT-SB HTL exhibit a linear spectrum followed by a semicircle bending towards the real axis (Figure 9d). In mass transit, two scenarios cause this behaviour. Ion diffusion occurs at a reactive barrier with a resistance several orders of magnitude lower than ion diffusion resistance. Ions react faster than they diffuse, preventing them from experiencing the interfacial boundary. In the second situation, ion diffusion to an adsorbing boundary results in a linear Nyquist plot and a semicircular path to the real axis, regardless of the reactiveness of the boundary. EIS recognize this as a finite transmissive threshold. The results indicate that the PVBT-SB acts as an adsorbing border for mobile $MA^+$ diffusion in the $MAPbI_3$ active layer. This theory was tested by separating the $MAPbI_3$/PVBT-SB interface with a PVBT-SO$_3$ layer (Figure 9b). The resulting EIS spectrum shows a linear Warburg component at low frequencies, indicating semi-infinite diffusion, similar to the PVBT-SO$_3$-only device. Ionic adsorption at the interface leads to ion accumulation and hinders responsiveness to opposite polarity fields.

In conclusion, the Randles circuit serves as a fundamental model in EIS for interpreting IS and modeling interfacial electrochemical reactions when there is semi-infinite linear diffusion of electroactive particles to flat electrodes. This circuit operates under the assumption that the rate of the faradaic reaction is governed by the diffusion of reactants to the electrode surface. While the model can be extended to represent more intricate electrochemical systems, it has limitations in accurately capturing the complexity of real systems due to its simplicity. Actual impedance spectra often exhibit more intricate patterns than what the Randles circuit can depict. Moreover, the assumption that the rate of the faradaic reaction is solely dictated by diffusion might oversimplify real electrochemical processes, leading to potential inaccuracies in data interpretation, particularly in systems with more intricate behaviors. Furthermore, the Randles circuit may not be suitable for systems with multiple overlapping processes, as it is primarily tailored for semi-infinite diffusion-controlled faradaic reactions to planar electrodes, thus restricting its applicability to more diverse electrochemical systems.

**Type 3. Transmission Line Model:**

It has been demonstrated that the transmission line model can be effectively utilized for the purpose of analyzing Dye Sensitized Solar Cells (DSSCs) and their variations, particularly those that make use of changeable colours like chlorophyll, anthocyanin, and beta-carotene. On the other hand, this model is most commonly seen in mesoporous $TiO_2$ arrangements for solid-state capacitors. This particular model's transmission pattern is characterized by a straight line that represents carrier transport, followed by an arc at lower frequencies that represents the coupling of capacitance with recombination resistance. This pattern is characterized by the fact that the arc occurs at lower frequencies.[34][80]

The Transmission Line Model (TLM) is a widely used equivalent circuit for perovskite solar cells, particularly for analyzing the electrical behavior of the device. The TLM is based on the concept of a series of transmission line segments, each representing a small section of the perovskite solar cell, where the electrical properties of the material are assumed to be constant.[81][82] The TLM is used to extract important parameters of PSC, such as the transport resistance for both electrons and holes, which can provide insights into the charge transport behavior of the device.[81] The TLM is also used to model the electrical behavior of the device under varying illumination intensity, which is important for understanding the mechanisms limiting energy harvesting in medium- and long-range wireless power transfer.[83]

However, the TLM has some limitations. For example, the TLM assumes that the electrical properties of the material are constant throughout the device, which may not be the case in real-world devices where the electrical properties may vary with the thickness of the perovskite layer.[82] Additionally, the TLM does not account for the effects of interfacial layers, such as the electron transport layer (ETL) and the hole transport layer (HTL), which can significantly impact the electrical behavior of the device.[82] Another limitation of the TLM is that it does not account for the effects of recombination, which can significantly impact the efficiency of the device.[82]Recombination is the process by which charge carriers (electrons and holes) combine and lose their energy, which can lead to a decrease in the efficiency of the device. Despite these limitations, the TLM is a useful tool for understanding the electrical behavior of perovskite solar cells and for extracting important parameters of the device. However, it is important to keep in mind the assumptions and limitations of the model when interpreting the results. It does not reveal PSCs' time-dependent behaviour and transitory processes. The model assumes uniform circuit elements and ignores nonlinear behaviour, reducing its accuracy. This model struggles to explain and extract essential phenomena like trap-assisted recombination and ion migration.

Figure 10 shows the transmission line model of PSCs, with $R_s$, $R_t$, $R_{BL}$, $R_r$, and $R_{HTL}$ representing series resistance, charge transfer resistance at the ETL/HTL perovskite interface, TiO$_2$ compact blocking layer, recombination resistance, and hole transport resistance. $C_{BL}$ and $C_\mu$ represent blocking layer and chemical capacitance, respectively. Victoria Gonzalez-Perdo [11]conducted a significant study in which she observed that the transmission line phenomena vanishes at greater biases (>0.5V) in PSCs, and instead, a Gerischer pattern occurs. This shift is indicative of a decreased degree of freedom in strong recombination, which ultimately results in a significantly shorter diffusion length ($L_d$) in comparison to the layer thickness (L).[11][84] Typically, the Gerischer pattern is seen in PSCs when the transport resistance ($R_{tr}$) is greater than the recombination resistance ($R_{rec}$). This causes the carrier diffusion length ($L_d$) to decrease to a point where it is less than the sample size or the layer thickness (L).[11]

For IS analysis by TLM, we took a study by Esmaiel Nouri [85] which is represented in Figure 11. The resistance ($R_s$) is typically consistent across all devices due to their structural similarities. However, a notable impression was observed in the charge-transfer resistance ($R_t$) values at the interfaces of the electron transport layer (ETL) or hole transport layer (HTL) with the perovskite layer between devices utilizing pristine NiO$_x$ and those employing optimized NiO$_x$−GO layers (specifically NiO$_x$−3GO). Nyquist plots, depicted in Figure 11, illustrate this difference under dark conditions and at the open-circuit voltage ($V_{OC}$), revealing internal charge transport and recombination within the PSC devices. The plots show a single semicircle, denoting the charge-transfer resistance ($R_t$) at the NiO$_x$ or NiO$_x$−GO and perovskite film interface, with the NiO$_x$−GO-based devices exhibiting lower $R_t$, implying faster charge transport at the interface of the HTL and perovskite. This observation aligns with conclusions drawn from photoluminescence measurements. Additionally, Nyquist plots under AM 1.5G illumination display high- and low-frequency semicircles, indicative of charge-transfer resistance ($R_t$), series resistance ($R_s$), dielectric contributions, and recombination resistance ($R_r$) at the photoactive layer. Fitting the plots with equivalent circuits shows that NiO$_x$−GO-based devices have smaller $R_t$ and larger $R_r$, indicating enhanced hole extraction and reduced charge recombination, leading to superior device performance compared to those based solely on pristine $NiO_x$. This suggests that while the presence of GO does not affect series resistance, it improves the structural organization of the HTL, facilitating charge transfer and impeding charge recombination. Furthermore, comparing the photoconversion efficiency over time of PSC devices using spiro-OMeTAD (as a reference standard HTL) with those employing $NiO_x$ and $NiO_x$–3GO, and either Au or carbon back contact electrodes, reveals superior performance of $NiO_x$–GO-based devices. These comparisons were conducted under controlled humidity and ambient air conditions, highlighting the importance of HTL composition on the photovoltaic performance of PSCs.

In summary, TLM is an EEC used to analyze the electrical behavior of PSCs. The TLM is based on the concept of a series of transmission line segments, each representing a small section of the device. The TLM is used to extract important parameters of the device, such as the transport resistance for both electrons and holes, and to model the electrical behavior of the device under varying illumination intensity. However, the TLM has some limitations, such as assuming constant electrical properties throughout the device, not accounting for interfacial layers, and not accounting for the effects of recombination.

**Type 4. Diffusion-Recombination Model:**

Additionally, the Diffusion-Recombination Model was developed with the express purpose of analyzing the behaviour of traps in disordered materials. This was accomplished by taking into consideration trapping factors and nonlinear recombination rates. The Transient Line Model was also developed for this purpose. Bisquert et al. have provided the amazing mathematical expressions for the newly improved model with regard to PSCs.[84] Reader can follow this article about calculation of impedance (Equation 40). A representation of the kinetics of minority charge carriers is included in this model. This Impedance representation can be expressed as follows:

$$z(\omega) = \left(\frac{r_{tr}.r_{rec}}{1+\frac{i\omega}{\omega_{rec}}}\right)^{1/2} cotanh\left[\left(\frac{r_{tr}}{r_{rec}}\right)^{1/2}\left(1+\frac{i\omega}{\omega_{rec}}\right)^{1/2}\right] \quad (40)$$

Within this context, the transport resistance and the recombination resistance are denoted by the symbols $r_{tr}$ and $r_{rec}$, respectively. The distinctive angular frequency of recombination is represented by the symbol, $\omega_{rec}$. On the other hand, $cotanh$ is a function that represents the diffusion length scenario and its impact on the resistance that is given through direct current.

In Figure 12, EEC of transmission line model for diffusion-recombination is represented. When the diffusion resistance is ignored in a high mobility semiconductor, there is no need to use a transmission line model since the transport characteristics can be omitted from the beginning. This is because the transport features will not be noticed in measurements because the overall resistance is so low. This is how a general transmission line shown in Figure 12a can be reduced to the lumped circuit shown in Figure 12c, which is the parallel connection of chemical capacitance and recombination resistance.

The effect of Warburg impedance is represented in Figure 8. Figure 13A illustrates the semi-infinite diffusion of chemical species, characterised by a single barrier at the electrode/electrolyte interface at $x = 0$. This behaviour is time-dependent and not steady state. In quiescent conditions, the diffusion layer extends to indefinite length ($x \to \infty$) as the concentration gradient decreases with time towards the main solution. The impedimetric response of Warburg impedance $Z_W$ in semi-infinite linear diffusion can be modelled using a transmission line with infinite length and a network of resistors and capacitors to represent diffusion resistance ($R_d$) and chemical capacitance ($C_d$) (Figure 13B).[88] In many electrochemical systems, the impedimetric response is not characterised by the semi-infinite linear diffusion model's Warburg impedance $Z_W$, but by a two-state model. The steady state diffusion area has a finite length, $x = L$, where L is the material length measured for diffusional impedance (Figure 13C). Figure 13C illustrates that one side of the material is permeable to ions engaged in the electrochemical process, while the other side can be either permeable (Figure 13D) or impermeable (Figure 13E). To account for the unique geometries of the electrochemical cell, it is more accurate to refer to the finite-diffusion region (not side) over time. After a time, span $t > t_d = \frac{2\pi}{\omega_d} = \frac{2\pi}{D/L}$ (D = diffusion coefficient) is either permeable or impervious to diffusing ions. The period t represents the time until the diffusing species cross the finite diffusion area. The concentration gradient is time-dependent until this point, resulting in semi-infinite diffusion behaviour. As ions diffuse throughout the diffusion region, the concentration gradient either remains constant or becomes zero.is either permeable or impervious to diffusing ions. The period t represents the time until the diffusing species cross the finite diffusion area. The concentration gradient is time-dependent until this point, resulting in semi-infinite diffusion behaviour. As ions diffuse throughout the diffusion region, the concentration gradient either remains constant or becomes zero.is either permeable or impervious to diffusing ions. The period t represents the time until the diffusing species cross the finite diffusion area. The concentration gradient is time-dependent until this point, resulting in semi-infinite diffusion behaviour. As ions diffuse throughout the diffusion region, the concentration gradient either remains constant or becomes zero.

Essentially, finite diffusion can be divided into two scenarios: (a) When the finite-diffusion zone is permeable to Electrochemical cells create a steady-state concentration gradient (dC/dx = constant) as species diffuse, resulting in a current flow. This is a porous or transmissive barrier. (b) Once the finite-diffusion area becomes impermeable to diffusing species, charge transfer is disrupted, and the concentration gradient of the species becomes zero (dC/dx = 0). This is a reflecting or impermeable barrier. For gaining much more understanding reader can refer to reference.[60]

Later investigations showed that ionic transport across the interface and within PSCs was important. PSCs were ionic-electronic mixed conductors.[15][86]Iodine ions made a big impact in the low-frequency (lf) region. This impaired low-frequency dielectric response and generated current-induced hysteresis. Thus, the diffusion-recombination model for PSCs, which only included electrons and holes, was inappropriate since it ignored ionic transport and build-up. PSC lf-IS spectra show that ETL and HTL selection greatly affects recombination resistance. Photo-generated electron-hole pair recombination reduces device efficiency. This resistance can be calculated using diffusion length or characteristic times from impedance measurements.[11]Two distinct temporal periods in the lf region suggest various recombination processes.[87] Additionally, the recombination method must account for intrinsic ion reorganisation in the perovskite layer as well as the lf area. Without HTL material (spiro-OMeTAD), transport resistance dominates in the high-frequency (hf) region. An interfacial recombination mechanism, not a bulk process, causes this phenomenon, emphasising charge separation in perovskites at hf.[16][30] However, selective interactions must be understood for complete insights.

In Summary, both the simplicity of the Diffusion-Recombination Model and the valuable insights it provides into charge carrier movement in semiconductor materials are advantages that the model offers. The fact that it does not consider ionic conductivity and does not consider the intricate recombination routes that are present in particular materials can, however, cause it to deviate from the experimental data. As a consequence of this, the model would not be able to accurately represent the behaviour of more complex systems, such as PSC, which feature mixed conductors and various recombination paths. As a result, they are less useful in situations like these. Within the context of addressing these difficulties, the Surface Polarisation has emerged as a useful model.

**Type 5. Surface Polarization Model:**

The Surface Polarization Model is a theoretical framework used to describe the dynamic hysteresis observed in perovskite solar cells. This model is based on the idea that the cell accumulates a large quantity of surface polarization due to the slow transient process, which is associated with the ion migration and charge accumulation at the interfaces.[12][51] The surface polarization model can explain the observed correlation between high apparent capacitances and hysteresis in current-voltage curves, which cannot be used to quantify charge accumulation. The respective resistances, however, are related to electron and hole injection, transport, and recombination in the solar cell, being mainly determined by the most limiting process. [51]

The Surface Polarization Model assumes of large electric and ionic charge accumulation at the surface and interfaces of the perovskite solar cell. This model can be used to understand the origin of apparent light-enhanced and negative capacitance in perovskite solar cells, which is due to the surface polarization effect.[51][89] The polarization field model is another theoretical framework used to describe the behavior of perovskite solar cells. This model is based on the Schottky barrier effect and depolarization field-driven photovoltaic effect, which can be used to understand the principles and advances of photoferroelectrics perovskite solar cells.[90]

Ravishankar et al. proposed a surface polarization model analysis of dynamic hysteresis with significant transient current to investigate polarization effects.[89] This study showed using fullerene contact instead of $TiO_2$ as a perovskite layer ETL eliminates PSC hysteresis effects at normal temperature. It was also shown that fullerene-based PSCs lack capacitance, which causes interface polarization, and that the interface charges electronically, creating transient current and hysteresis. These findings match literature.[12]  This study's capacitance-frequency plots show that ionic polarization is low frequency at the perovskite interface while dielectric polarization is high frequency in the bulk.

Another understanding of the surface polarization effect and the behaviour of inductive loops in PSCs was achieved by Ghahremanirad and colleagues through the development of a dynamic hysteresis model that included the EIS response.[51] The model takes into account a kinetic relaxation period for ion displacement, which leads to the formation of separate impedance arcs that have their own set of underlying distinctive characteristics. Important kinetic information regarding the interfaces of PSCs, such as delayed surface voltage and its strict dependence on kinetic relaxation time, can be obtained from the inductive loops that have been discovered. These loops indicate positive and negative capacitance.

In order to highlight the dynamic hysteresis behaviour of PSCs, Figure 14 depicts the EEC of surface-polarization model. This model incorporates the electrical elements that are connected with the motions of ions. In this context, the symbol $C_d$ represents the dielectric capacitance that was created in order to specifically account for dielectric bulk/surface phenomena experienced at higher frequencies. Series resistance, recombination resistance, and series resistance with inductive effects (L) are the elements that correspond to the series resistance, recombination resistance, and series resistance, respectively. In addition, the symbol $R_c$ represents the series resistance in conjunction with the capacitance $C_1$, and the product of these two variables, denoted as ($R_c/C_1$), may be used to get the time constant value, which governs the kinetics process within the PSCs. Reader can refer to [89], [91] for calculation of impedance of EEC represented in Figure 15.

$$Z(\omega) = R_s + \left[ j\omega C_g + \frac{1}{R_b} + \frac{1}{R_a + j\omega L_a} + \frac{1}{R_x + \frac{1}{j\omega C_x}} \right] \quad (41)$$

The following is possible interpretation of the circuit from a physical standpoint: the parallel combination of $R_{rec}$ and $C_d$ creates a typical circuit for solar cells, which is controlled by recombination. An additional circuit, denoted as $R_C$ - $C_1$, is responsible for the charging and discharging of the surface accumulation capacitance. This process is governed by the kinetic relaxation constant, $\tau$-$_{kin}$. It is a consequence of charge buildup and extraction at the surface. The non-standard process $R_L$ - L, which is a prominent characteristic of the model, is incorporated in order to account for the temporal delay in surface charging that is caused by the motions of ions. As a result of the presence of an inductive line in the analogous circuit, this delay is controlled by the, V-I. This variable determines the pace at which the interface reaches equilibrium, which in turn leads to the presence of the inductive line.

For better understanding of IS we took a latest study of 2024 by Balaguera [91] which is represented in Figure 15. in this they interpret the importance of crucial representation of the frequency-dependent capacitance in the Bode plot representation (see Figure 15b). This rephrasing of the impedance data traces a transition between two capacitive plateaus, producing negative values with undershoots in the middle-frequency region. These undershoots are related with the inductive loop that was detected in the impedance. These relaxation processes, which underlying physical properties, are represented by the various elements of the equivalent circuit shown in Figure 15c. $R_s$ is the parasitic series resistance, $R_b$ and $R_a$ are the two branches of the recombination resistance, and $R_x$ is the resistance for ionic relaxation. The dielectric capacitance is denoted by the symbol $C_g$, the chemical inductance is denoted by $L_a$, and the interfacial charging capacitance is the symbol $C_x$. Due to the fact that it makes use of the vast array of ambiguous perovskite patterns discovered through IS in a manner that is both universal and physiologically motivated, the EEC depicted in Figure 15c is the most successful equivalent circuit for the impedance response. Impedance is represented by Equation 33, which shows the complexity of EEC (Figure 15c). Readers can refer the reference[91] for much deep understanding.

The surface polarization model was also utilized in research such as [31][89][92] in order to gain a better understanding of the data that was collected during the creation of positive/negative inductive loops. These investigations shed information on the role that RL plays in determining the cause of the delay in surface charge that is brought about by the passage of ions. The capacitance $C_1$ is associated to the low-frequency area, while the dielectric capacitance $C_d$ is reflected in the high-frequency portion of the spectrum. To obtain a comprehensive understanding of this phenomena, however, it is required to have specialized insights into the kinetic process, interfacial chemistry, and the underlying physical components that influence device performance. This is for the purpose of acquiring a comprehensive understanding of this phenomenon.

In summary, the Surface Polarization Model is a theoretical framework used to describe the dynamic hysteresis observed in PSC. This model is based on the idea that the cell accumulates a large quantity of surface polarization due to the slow transient process, which is associated with the ion migration and charge accumulation at the interfaces. The polarization field model is another theoretical framework used to describe the behavior of perovskite solar cells, which is based on the Schottky barrier effect and depolarization field-driven photovoltaic effect.

**Type 6. Neuron style Model:**

In 2022, Gonzales et al. developed the 'Neuron Style Model' to gain deeper insights into the hysteresis behavior of J-V and EIS response in Metal Halide PSCs and to understand the formation of negative capacitance (inductive loop) at high voltage.[53] This model is essential for better understanding and optimizing the performance of perovskite solar cells. They proposed a Neuron Style Model on the basis of differential equations which is valid for any type of perturbation. According to this model, total current flowing through the device is divided into several components. Firstly, a portion of the total current ($I_{tot}$) corresponds to displacement current, which charges the geometrical capacitance (Cg) of the perovskite material. The remaining current ($I_v$) is then subdivided into three branches: Charging the surface capacitance ($C_1$), Extracting rapid current ($I_{cat}$) at the contact, and extracting current slowly via an ion-modulated current ($i_d$). This systematic division allows for a comprehensive understanding of the current flow dynamics within the device, highlighting the contribution of each component towards its overall operation and performance. These equations define this model:

$$I_{tot} = C_g \frac{du}{dt} + I_v(v) \qquad (42)$$

$$I_v(v) = \frac{dQ_s(w)}{dt} + I_c(\omega) + i_d \qquad (43)$$

$$\tau_d \frac{di_d}{dt} = I_d(w) - i_d \qquad (44)$$

Here $I_{tot}$, voltage (u) are external variables. $I_v(v)$ and $I_c(\omega)$ are instantaneous functions of the respective voltage, $I_d$ is variable current with characteristic time $\tau_d$ due to ionic effects. $Q_s(w)$ is surface charge function and the stationary value of the delayed current denoted by $I_d(w)$. Impedance in Laplace form (where $s = i\omega$) of EEC which are represented in Figure 12(d) can be represented by

$$Z(s) = \left[ C_g s + \cfrac{1}{R_3 + \cfrac{1}{C_1 s + \cfrac{1}{\frac{1}{R_1} + \frac{1}{R_a + L_a s}}}} \right] \qquad (45)$$

A chemical inductor arrangement that describes delayed surface-controlled recombination is the branch created by $R_a$ and $L_a$. The transition from capacitance to inductive behaviour is often connected with the design and configuration of the parts that make up the circuit. The slower ionic displacement has an effect on the significantly faster recombination of electronic current or the transfer of that current towards contacts. As a result of the voltage-gated ion channel in neuron membranes that is represented this behaviour is frequently associated with the presence of a chemical inductor.[93] The voltage-gated electronic current that is seen at the surface of solar cells can be better understood with the help of this model.

In this they analyzed the IS into three voltage regions (low, medium, and high). The components of the circuit that are utilized at the various voltage levels (low, transition, and high) that are applied to the MAPBr solar cell are illustrated in Figure 16. The observation of a double RC feature can be seen in Figure 16 (b-c) when the voltages are lower (as shown in Figure 16), which is caused by the inductor having a high resistance. A low-frequency arc is replaced by an incipient inductive feature in the intermediate/transition zone, which is denoted by the EEC in Figure 16 (b). This can be seen in Figure 16 (e-f). The capacitive arc that is connected with $C_1$ is not observed at higher voltages (Fig. 16 g), and it is replaced by a chemical inductor, which therefore results in a newly observed inductive arc effect (Fig. 16 (h-i)).

The advantage of Neuron Style Model for solar cells, specifically focusing on halide perovskite cells transitioning from capacitive to inductive behavior at varying voltages. their model accurately replicates experimental features like impedance spectroscopy and hysteresis in current-voltage curves under different sweep rates. They found that capacitive excess current scales with the scan rate, while inductive current inversely scales with it. This model enables precise control over cell kinetics, predicting hysteresis characteristics based on impedance measurements. Additionally, it allows correlation of hysteresis with material properties via impedance parameters. The model explains the transition of the low-frequency capacitor to a chemical inductor, regulated by certain time constants, preserving regular hysteresis across the voltage range without negative capacitance. In spite of these limitations, the Neuron Style Model continues to be an invaluable technique for the purpose of investigating and optimizing MHPSCs. It has also been utilized in a number of different sources of literature, such as.[46], [58], [94], [95]

In this section, we conclude the physical modelling of EEC for PSCs, emphasizing the representation of progress over time in understanding their dynamic behavior. By incorporating essential equivalent circuit elements like resistors, capacitors, and inductors, we can effectively simulate and analyze the temporal changes in the electrical response of PSCs. This approach offers insights into transient phenomena, charge carrier dynamics, and time-dependent processes within the device, vital for optimizing performance and stability. As the field evolves, continuous innovation in circuit modeling is crucial, pushing boundaries to capture the intricate dynamics of PSCs. The commitment to refining and customizing equivalent circuits is paramount, as it enables a deeper understanding of underlying physical phenomena, ultimately leading to improved performance and advancement of PSCs.

## 4. Suggestions, conclusion, and outlook

Electrochemical systems, like typical PSC, involve complex processes such as charge transfer and diffusion of charged species. Understanding these processes is crucial for analyzing PSCs, as the interaction between different layers and materials adds to the complexity. Impedance spectra of PSC devices show a mix of processes, making interpretation challenging without considering kinetic parameters. Factors like ion migration, electrode porosity, and non-homogeneity must be accounted for in EEC models based on physical origins, enabling precise analysis and insights into PSC behavior.

Physically motivated EECs help uncover the origins of resistances and capacitances in electrochemical systems, eliminating ambiguity through auxiliary techniques or clear argument chains. Before conducting EIS, theoretical understanding of impedance origins is essential to avoid cognitive bias in interpretation. External parameters like temperature and bias potentials aid in exploring interface impedances, while attention to system non-stationarity prevents impedance response distortions. The multistage intercalation mechanism in PSCs illustrates the superiority of EECs with clear physical meaning, enabling precise interpretation of battery system performance and facilitating the design of new electrode structures.

Research in the emerging field of PSCs is highly interdisciplinary, drawing expertise from diverse fields including chemistry, solid-state physics, material science, and electrical engineering. However, beginners often encounter challenges in grasping how IS effectively studies new PSC materials and devices due to its broad scope. In this review, we emphasize the potency of EIS in characterizing interfaces and understanding dynamic processes within PSCs. By analyzing impedance spectra, crucial insights into the electrical and electrochemical properties of PSCs are collected, particularly concerning various interfaces such as ITO/FTO contact surfaces, transport layers, and electrode-perovskite interfaces. IS applications have greatly contributed to comprehending charge transport, recombination processes, and interface effects on device performance in PSCs. However, modelling equivalent circuits remains a challenge. This paper presents advancements in equivalent circuit modelling, facilitating the extraction of parameters crucial for understanding electrochemical behaviour. The review aims to provide researchers with a comprehensive toolkit for investigating and analyzing novel photovoltaic materials using impedance spectroscopy, bridging gaps in literature by offering insights into equivalent circuit elements not readily available elsewhere.

While significant progress has been made in applying IS to PSCs research, there are still several areas that offer potential for future exploration and advancement. The prospects of IS in the field of PSC are promising and hold great potential. IS can help in improving and characterize interface optimization, involving characterization, defect detection, and stability assessment, remains a focal point for enhancing the effectiveness of solar panels. Advanced circuit models play a pivotal role in accurately analyzing impedance spectra, considering factors like double-layer capacitance and interfacial reactions. Stability and degradation Studies utilize IS to monitor long-term performance, identifying degradation mechanisms and enhancing overall stability. Exploration of new materials and architectures through IS provides crucial insights into improving solar panel efficiency. Non-destructive testing and monitoring, coupled with IS, ensures continuous evaluation and reliability of solar systems. The pursuit of low-frequency response analysis and integration of ML aims to overcome challenges and improve predictive capabilities. Additionally, device performance optimization and integration with other techniques underscore the importance of IS in enhancing solar panel efficiency and understanding materials comprehensively. Further exploration and integration of these aspects promise to advance the field of solar energy research towards greater efficiency and stability.

Overall, the future of IS in PSCs research is bright, and research lies in the continued refinement of measurement techniques, the development of advanced circuit models, and the integration of EIS with other characterization methods. It offers opportunities for improved material understanding of device optimization, reliability assessment, and the development of next-generation photovoltaic technologies. By further exploring the electrochemical responses of interfaces and materials, EIS can contribute to the development of more efficient, stable, and reliable PSCs. Continued research and technological advancements in EIS will contribute to the advancement and widespread adoption of efficient and sustainable solar energy solutions.

**Statements & Declarations**

**Competing Interests:** The authors declare that they have no conflict of interest.

**Author Contributions:** All the authors contributed a review paper. Rajat Kumar Goyal designed it. M. Chandrasekhar have given valuable comments to rectify problems. Rajat Kumar Goyal wrote the first and final draft of the manuscript. All the authors commented on the previous version, read, and approved the final manuscript.

**Figure captions**

**Figure 1:** Electrochemical impedance spectroscopy (EIS) is a way to figure out things by causing a disturbance in a balanced system and studying how it responds. This involves using an external signal to disturb the system and then looking at the electrical data that comes out of it. An equivalent electrical circuit (EEC) model used to represent different processes happening in the system.

**Figure 2:** (a) Illustration of various types of EEC elements, (b) Various types of resistance [39], (c) various types of capacitances [40] and (d) Inductors used in PSCs [41].

**Figure 3:** Nyquist, Bode magnitude and phase angle plots of some basic model circuits simulated by Z-view for understanding.[60]

**Figure 4:** Nyquist, Bode magnitude and phase angle plots of some basic model circuits simulated by Z-view for understanding.[60]

**Figure 5:** Illustration of Impedance Spectra Validation. [60][72]

**Figure 6:** An illustration of various extractable underlying phenomena in impedance spectra across different frequency and time range.[37]

**Figure 7:** (a) IS on devices with mp-$TiO_2$ and c-$TiO_2$/m-$TiO_2$ (b) Voight Circuit (widely used circuit), utilized for circuit fitting (c) Corresponding their total impedance. [75]

**Figure 8:** (a) Randles Circuit, (b) conversion at high frequency and (c) at low frequency (d) simulated IS by EIS analyzer according to circuit (b) (e) according to circuit (c) (f) full spectra of Randles circuit according to circuit.

**Figure 9:** Exploration of ion transport at interface (a) & (b) device used for studying properties (c) circuit used (d) corresponding impedance spectra. [79]

**Figure 10:** Transmission model of PSCs.[85]

**Figure 11**: Nyquist plots EEC of the $NiO_x$- and $NiO_x$−GO based PSC devices, these plots and circuits were measured under two different illumination conditions: (a) dark and (b) AM 1.5G illumination. The bias voltage used was $V_{OC}$. The fitted curves are represented by the solid lines.[85]

**Figure 12:** (a) EEC of transmission line model for diffusion-recombination (b) same when $R_{tr}$ is negligible (c) case (b) can be written simplified form.[44] [84]

**Figure 13:** (A) The semi-infinite regime of an electrochemical cell, (B) the transmission line (TL) depiction of the semi-infinite regime, (C) finite boundary diffusion at t < $t_d$ spans, (D) transmissive, and (E) reflective boundary at t > $t_d$. [60]

**Figure 14:** Equivalent circuit of Surface Polarization Model.[89]

**Figure 15:** Latest study by Balaguera (2024) (a) device structure corresponding IS (b) capacitance frequency plot (c) corresponding EEC. [91]

**Figure 16:** IS examination of MAPBr solar cell under a varying dark condition and applied voltages ($V_a$). At low $V_a$ (<1.1 V), capacitive behaviour is found in circuits (b, c) that match to circuits (a). A transition zone may be observed at a voltage range of 1.1 V to 1.2 V (e, f) associated with circuit (d). Circuit (g) exhibits inductive behaviour when the applied voltage is high (more than 1.2 V) (h, i).[53]

# Figures

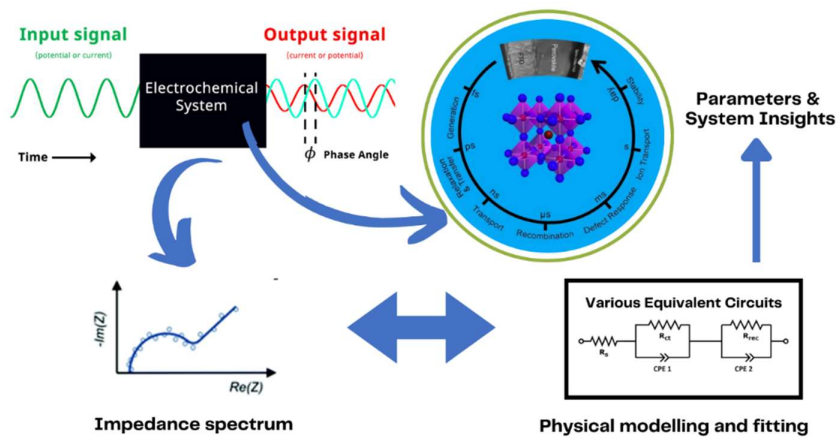

**Figure 4:** Electrochemical impedance spectroscopy (EIS) is a way to figure out things by causing a disturbance in a balanced system and studying how it responds. This involves using an external signal to disturb the system and then looking at the electrical data that comes out of it. An equivalent electrical circuit (EEC) model used to represent different processes happening in the system.

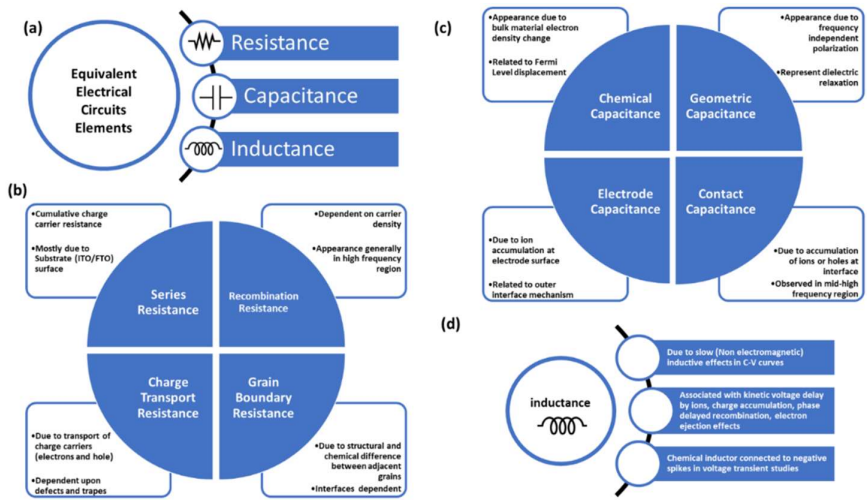

**Figure 5:** (a) Illustration of various types of EEC elements, (b) Various types of resistance [39], (c) various types of capacitances [40] and (d) Inductors used in PSCs. [41]

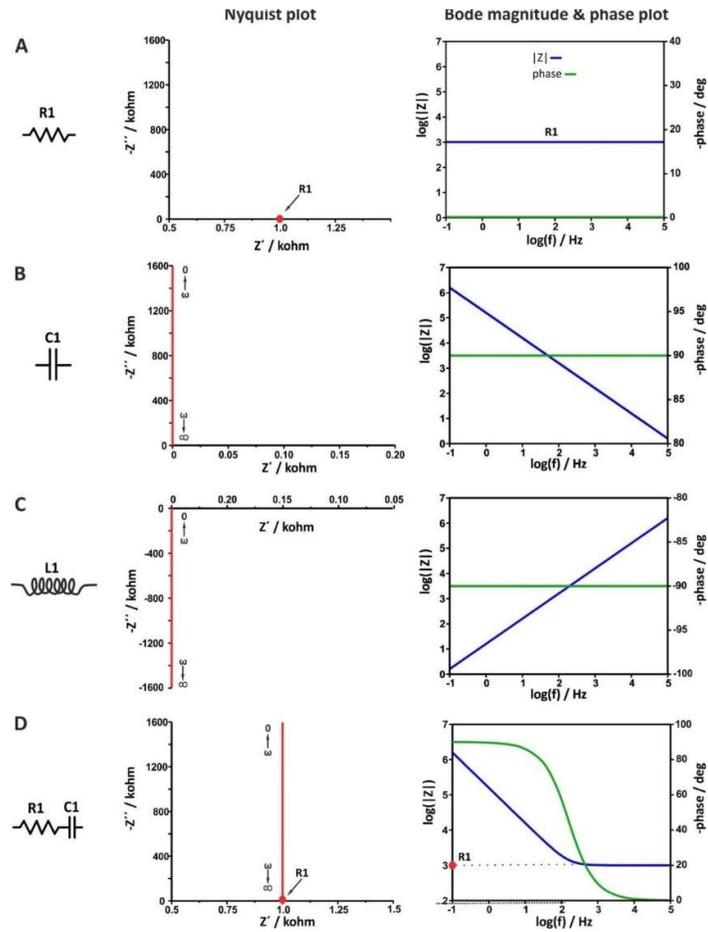

**Figure 3:** Nyquist, Bode magnitude and phase angle plots of some basic model circuits simulated by Z-view for understanding.[60]

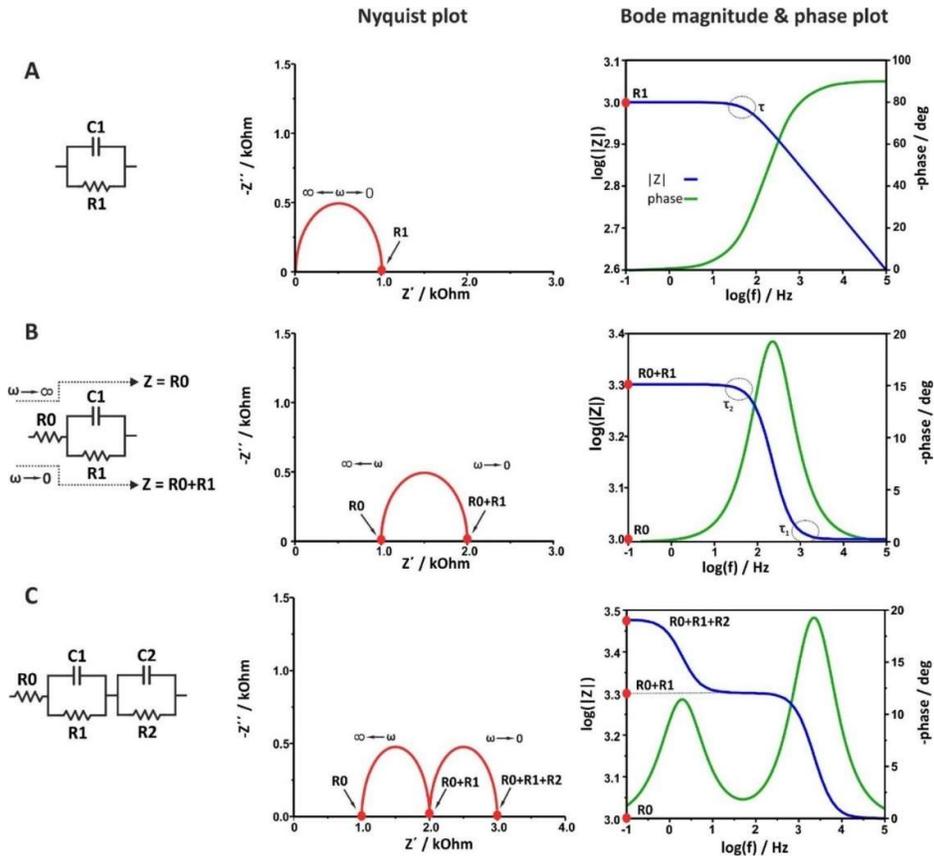

**Figure 4:** Nyquist, Bode magnitude and phase angle plots of some basic model circuits simulated by Z-view for understanding.[60]

**Figure 5:** Illustration of Impedance Spectra Validation. [60][72]

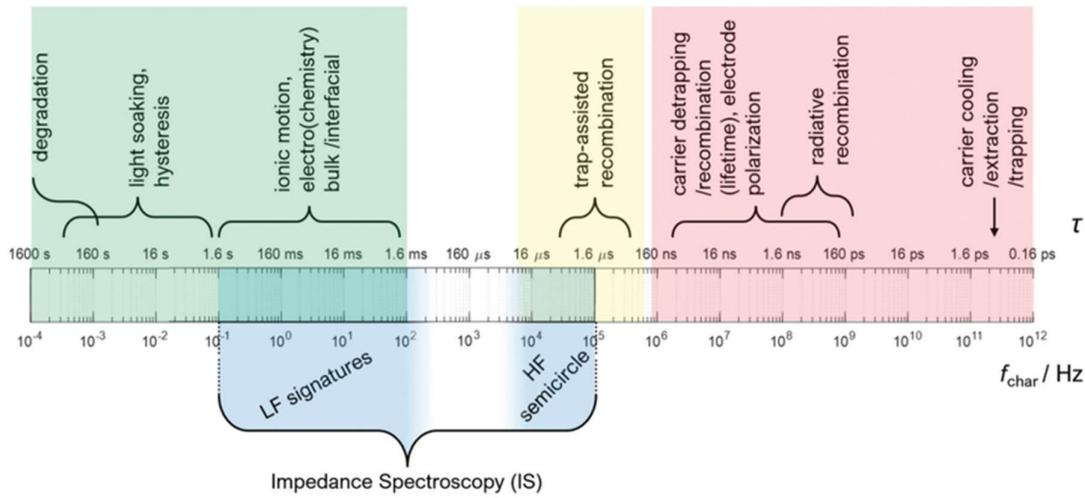

**Figure 6:** An illustration of various extractable underlying phenomena in impedance spectra across different frequency and time range.[37]

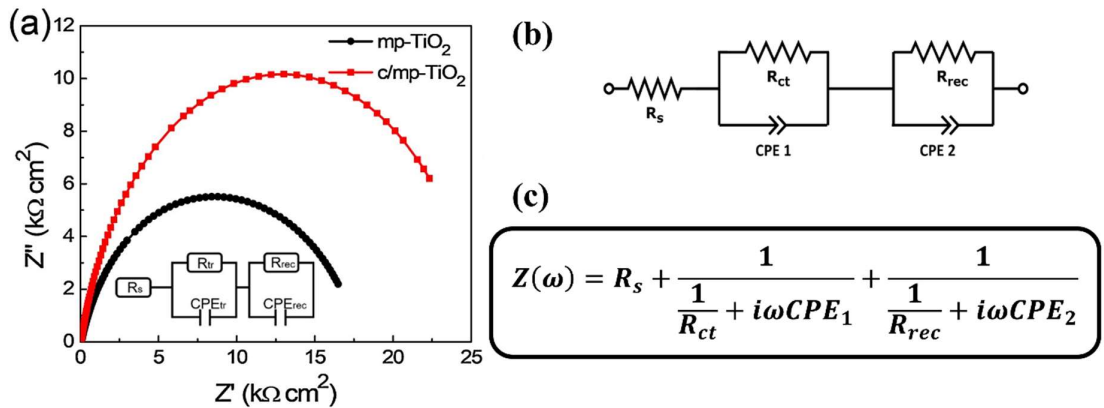

**Figure 7:** (a) IS on devices with mp-TiO$_2$ and c-TiO$_2$/m-TiO$_2$ (b) Voight Circuit (widely used circuit), utilized for circuit fitting (c) Corresponding their total impedance. [75]

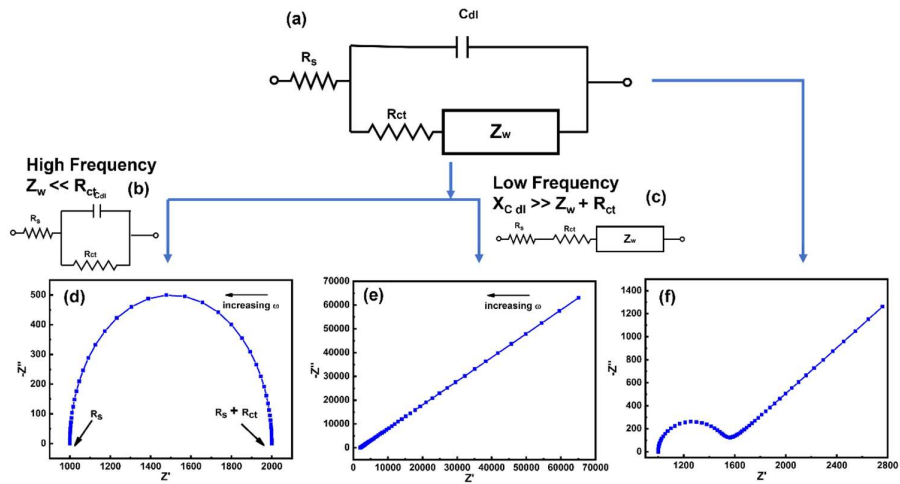

**Figure 8:** (a) Randles Circuit, (b) conversion at high frequency and (c) at low frequency (d) simulated IS by EIS analyzer according to circuit (b) (e) according to circuit (c) (f) full spectra of Randles circuit according to circuit.

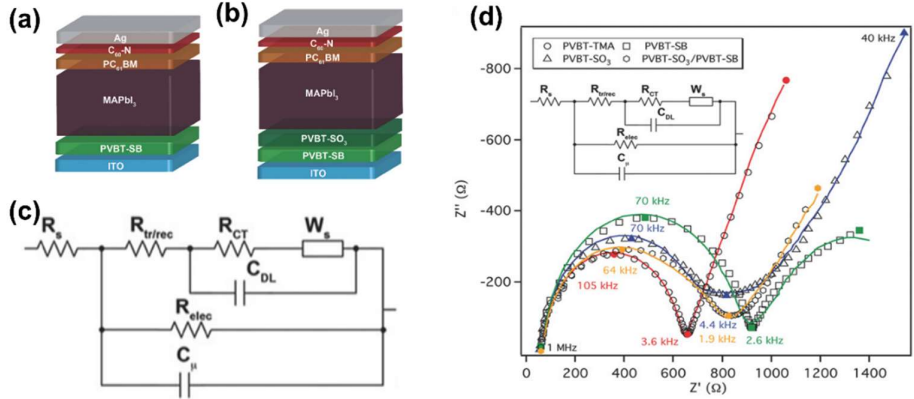

**Figure 9:** Exploration of ion transport at interface (a) & (b) device used for studying properties (c) circuit used (d) corresponding impedance spectra. [79]

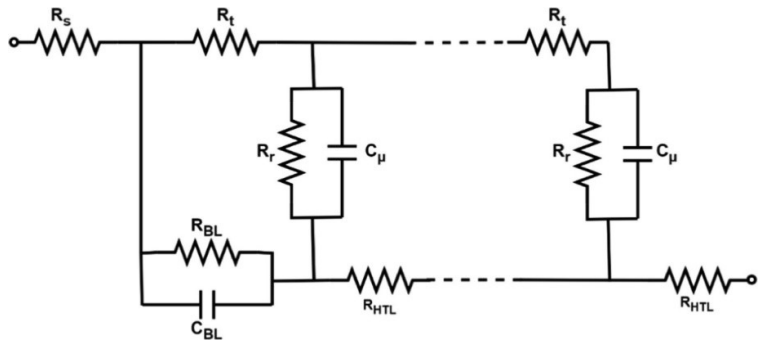

**Figure 10:** Transmission model of PSCs.[85]

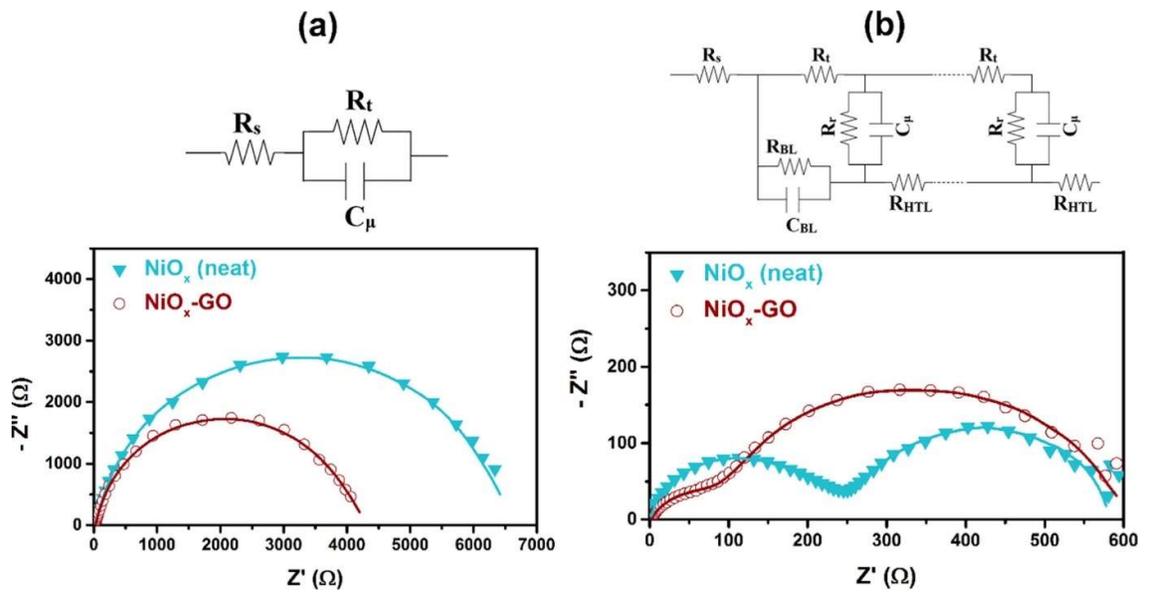

**Figure 11**: Nyquist plots EEC of the $NiO_x$- and $NiO_x$−GO based PSC devices, these plots and circuits were measured under two different illumination conditions: (a) dark and (b) AM 1.5G illumination. The bias voltage used was $V_{OC}$. The fitted curves are represented by the solid lines.[85]

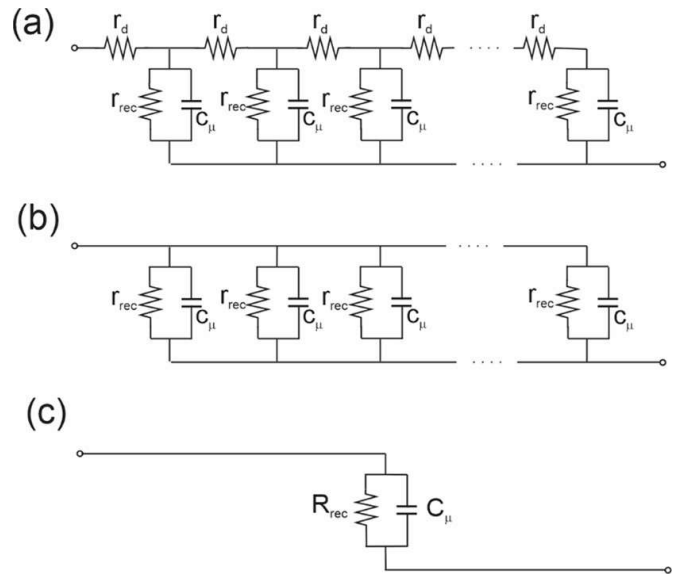

**Figure 12:** (a) EEC of transmission line model for diffusion-recombination (b) same when $R_{tr}$ is negligible (c) case (b) can be written simplified form.[44] [84]

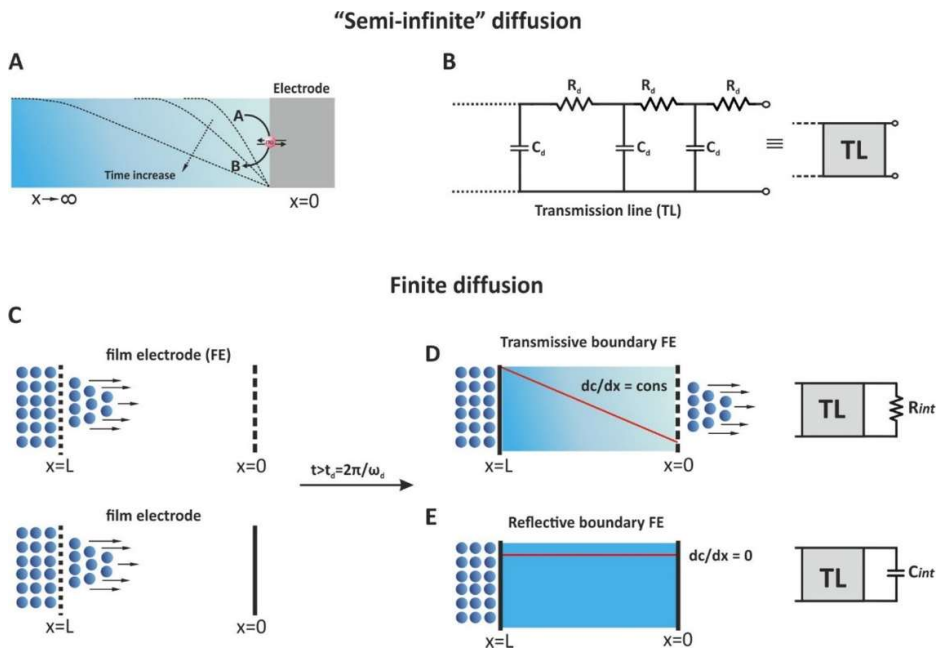

**Figure 16:** (A) The semi-infinite regime of an electrochemical cell, (B) the transmission line (TL) depiction of the semi-infinite regime, (C) finite boundary diffusion at $t < t_d$ spans, (D) transmissive, and (E) reflective boundary at $t > t_d$. [60]

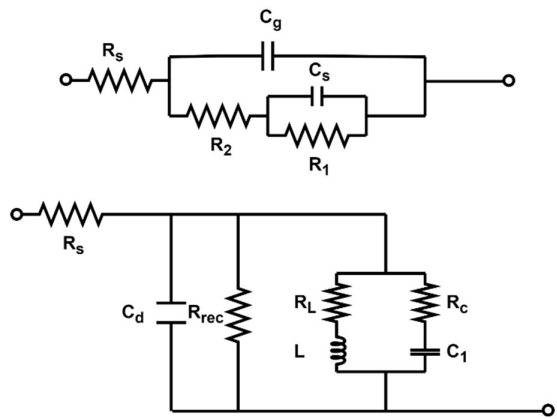

**Figure 14:** Equivalent circuit of Surface Polarization Model.[89]

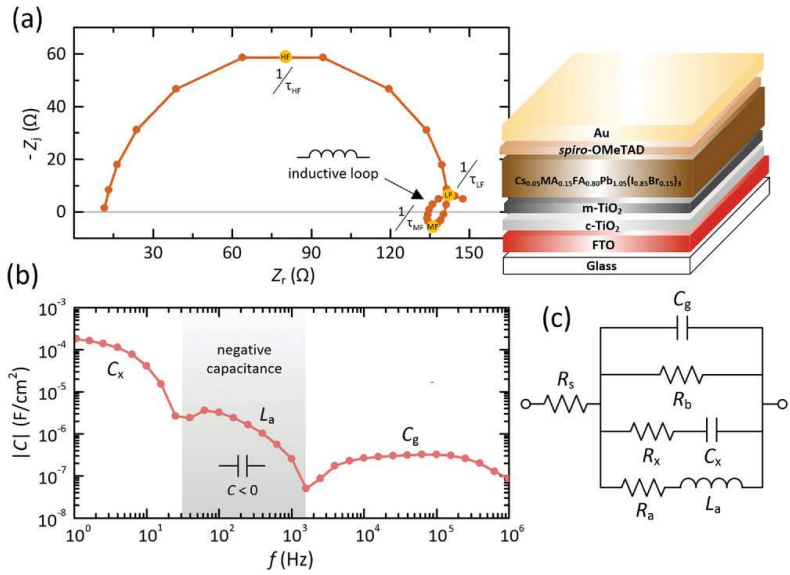

**Figure 15:** Latest study by Balaguera (2024) (a) device structure corresponding IS (b) capacitance frequency plot (c) corresponding EEC. [91]

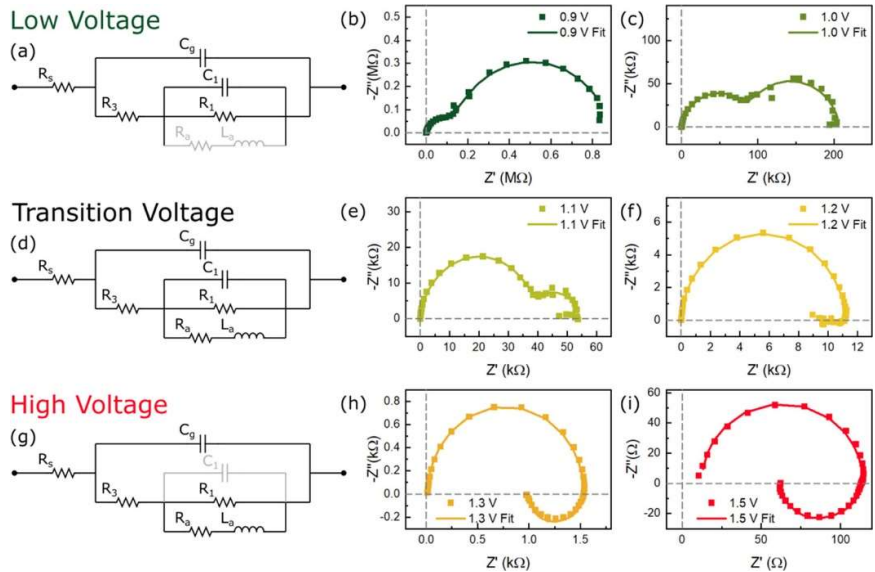

**Figure 16:** IS examination of MAPBr solar cell under a varying dark condition and applied voltages ($V_a$). At low $V_a$ (<1.1 V), capacitive behaviour is found in circuits (b, c) that match to circuits (a). A transition zone may be observed at a voltage range of 1.1 V to 1.2 V (e, f) associated with circuit (d). Circuit (g) exhibits inductive behaviour when the applied voltage is high (more than 1.2 V) (h, i).[53]